\documentclass[final,1p,times]{elsarticle}

\usepackage{amssymb}
\usepackage{amsthm}
\usepackage{amsmath}
\usepackage{float}
\usepackage{subfig}
\usepackage{lipsum}
\usepackage{graphicx}
\usepackage{dcolumn}
\usepackage{bm}
\usepackage[hidelinks]{hyperref}
\usepackage{xcolor}
\usepackage{physics}
\usepackage[normalem]{ulem}
\makeatletter
\renewcommand{\p@subfigure}{\thefigure--}
\makeatother

\definecolor{red}{rgb}{1, 0, 0}
\definecolor{bluel}{rgb}{0, 1, 1}
\definecolor{blues}{rgb}{0, 0, 1}
\definecolor{green}{rgb}{0, 1, 0}
\definecolor{orange}{rgb}{1, 0.5, 0}
\definecolor{gdark}{rgb}{0, 0.4, 0}
\hypersetup{
    colorlinks,
    linkcolor={magenta!50!black},
    citecolor={blue!50!black},
    urlcolor={blue!80!black}
}
\usepackage{algorithm}
\usepackage[noend]{algpseudocode}

\newcommand{\bn}[1]{\mbox{\boldmath $#1$}}

\def\mb{\mbox}

\def\bm{\boldmath}

\def\ni{\noindent}

\journal{Journal of Computational Physics}

\begin{document}

\begin{frontmatter}

\author[label1]{E. Nieva-P\'erez}
\cortext[cor1]{Corresponding Author}
\address[label1]{Departamento de F\'isica y Matem\'aticas, Universidad Iberoamericana, CDMX, M\'exico}
\address[label2]{Facultad de F\'{\i}sica, Universidad de La Habana, La Habana, Cuba}
\address[label3]{Grupo de Materia Condensada, Instituto de F\'{\i}sica, Fac.de Ciencias Exactas y Naturales, Universidad de Antioquia, Medell\'{\i}n, Colombia}
\address[label4]{Departamento de Matem\'atica, Facultad de Ingenier\'{\i}a y Tecnolog\'{\i}as, Universidad Cat\'{o}lica de Uruguay, Uruguay}
\author[label1]{E. A. Mendoza-\'Alvarez}
\author[label1,label2]{L. Diago-Cisneros\corref{cor1}}
 \ead{ldiago@fisica.uh.cu}
\author[label3]{C. A. Duque}
\author[label4]{J. J. Flores-Godoy}
\author[label1]{G. Fern\'andez-Anaya}

\title{A bandmixing treatment for multiband-coupled systems \emph{via} nonlinear-eigenvalue scenario.}

\begin{abstract}
We present a numeric-computational procedure to deal with the intricate bandmixing phenomenology in the framework of the quadratic eigenvalue problem (QEP), which is derived from a physical system described by $N$-coupled components Sturm-Liouville matrix boundary-equation. The modeling retrieves the generalized Schur decomposition and the \textit{root-locus-like} techniques to describe the dynamics of heavy holes (\textit{hh}), light holes (\textit{lh}) and spin-split holes (\textit{sh}) in layered semiconductor heterostructures. By exercising the extended $(N = 6)$ Kohn L\"uttinger model, our approach successfully overcomes the medium-intensity regime for quasi-particle coupling of previous theoretical studies. As a bonus, the sufficient conditions for a generalized QEP have been refined. The \textit{sh}-related \textit{off}-diagonal elements in the QEP mass-matrix, becomes a competitor of the bandmixing parameter, leading the \textit{hh-sh} and \textit{lh-sh} spectral distribution to change, then they can not be disregarded or zeroed, as was assumed in previous theoretical studies. Thereby, we unambiguously predict that several of the new features detected for \textit{hh-lh-sh} spectral properties and propagating modes, become directly influenced by the metamorphosis of the effective \emph{band-offset} scattering profile due sub-bandmixing effects strongly modulated with the assistance of \textit{sh}, even at low-intensity mixing regime.
\end{abstract}


\end{frontmatter}

\section{Introduction}
\label{sec:Intro}

Commonly, the analysis of dynamic elementary excitations lead to treat with \emph{eigen}-systems, whose solutions characterize fundamental physical quantities of a wide variety of areas. One of this kind of systems within a nonlinear scenario, is the so-called quadratic eigenvalue problem (QEP) \cite{lancaster,tisseur}. The QEP remains relevant for different scientific topics and its convincing advantages \cite{Diago2006,mendoza,job,memsqep}, motivate us to apply it for the present study of multiband-multicomponent systems (MMS). The transport properties of spin-charge carriers in semiconductor heterostructures, are nowadays receiving a renewed interest toward technological applications such as electronic and optoelectronic devices \cite{Voss,memsarches,xiangyu}. Most of the prior studies focuses the electronic case, because there is a better theoretical and practical foundation. Nevertheless, the study of holes has been increasing since it has been proved \cite{Myronov2015,mendez,Diago2006} that they have a crucial influence on the threshold response of devices based-on semiconducting layered heterostructures, when such quasi-particles are involved as slower spin-charge carriers. Therefore, the challenge of new theoretical frameworks for a better description of holes is still alive.

Usually the examination of bandmixing-free or weakly-coupled MMS can be done by one-dimensional (1D) Wannier functions \cite{wannier,Rosner2015}, transfer matrices \cite{Perez2004}, among other techniques \cite{multiaprox}. Nonetheless, for strongly-coupled MMS a non-parabolic bandmixing arises and then, conventional 1D or even uncoupled (weakly-coupled) $N$-component approximations are no longer valid. To address this type of systems, several approaches are widely used, and just to mention a few of them we remark: the tight-binding approximation  \cite{tbinding,ihn} for the valence band (VB) and the \mb{\bm $\vec{k} \cdot \vec{p}$} approximation \cite{Kane,multiaprox} for the bandmixing between the VB and the conduction band (CB). These models are grounded on multicomponent ($N\times N$) effective Hamiltonians, whose \emph{eigen}-solutions can be obtained by the envelope function approximation (EFA). Currently, there is a remarkable interest in strongly-coupled MMS for further potential development of technological appliances and devices \cite{xiangyu,winklerspin,ekbote,doty}.

We present in Section \ref{sec:Teo} a  theoretical treatment for bandmixed MMS \emph{via} the nonlinear-eigenvalue scenario of the QEP.  Our model, intrinsically avoids some inconsistencies in the determination of the transmission coefficients \cite{chao,wessel,kumar} and also the use of reduced Hilbert spaces \cite{kumar,dsanchez}. Therefore, we prevent the loss of physical-system's information, the applications of arbitrary factors to normalize the \emph{eigen}-spinors, and the lack of several physical symmetries that fulfills solely within the full Hilbert space \cite{diagoscrip}.

There are two targets focused in the present study. Firstly, to obtain the \emph{eigen}-spinors of the QEP that underlies an $N$-coupled components Generalized Sturm-Liouville (GSL) matrix boundary-equation \cite{Pernas2015,malik,moli}. By modeling coupled MMS, we recall the generalized Schur decomposition, the simultaneous triangularization and \textit{root-locus-like} techniques to describe the dynamics of heavy holes (\textit{hh}), light holes (\textit{lh}) and spin-split holes (\textit{sh}) in layered semiconductor heterostructures. Secondly, we would like to compare our simulations to previous theoretical calculations, and further predict whether or not several observed features for the \textit{hh-lh-sh} spectral properties and propagating modes, are directly influenced by  \textit{sh} sub-bandmixing effects.

For the sake of taking advantages from the EFA framework \cite{burt,foreman,multiaprox} we are working with, a proper manipulation of the basis to expand the \emph{eigen}-spinors is then required. For that, it is enough to choose correctly a second order differential systems for the envisioned $N$-coupled components bands \cite{guillermo,multiaprox}. There are plenty of examples of the correctness of such procedure \cite{mendoza,KL,ekbote,ahn,mkane,kanemod}.

The remaining part of this paper is organized as follows: In Section \ref{sec:Teo} will be addressed the theoretical outlines for a description of $N$-coupled MMS. The analytical and numerical treatment is presented in this section. Next, in Section \ref{sec:Sim} we will discuss and show the obtained results by applying the \textit{root-locus-type} technique for observing the \textit{hh-lh-sh} spectral distribution, as well as the evolution of the effective \emph{band-offset} scattering profile due sub-bandmixing effects. Finally, in Section \ref{sec:Con} our concluding remarks will be presented.

\section{Theoretical Approach}
 \label{sec:Teo}
 \subsection{Extended Kohn L\"uttinger Hamiltonian}
  The extended $(N = 6)$ Kohn L\"uttinger (KL) model \cite{ahn}, accurately describes \textit{hh-lh-sh} subbands, which are degenerated at the top of the VB in the absence of bandmixing and spin-orbit coupling. For the usual $B=0$ case, the effective Hamiltonian $H_{\mb{\tiny KL}(6 \times 6)}$  corresponding to $\Gamma_{8v}$ and $\Gamma_{7v}$ VB states in the basis $|\frac{3}{2},\frac{3}{2}\rangle$, $|\frac{3}{2},\frac{-1}{2}\rangle$, $|\frac{1}{2},\frac{-1}{2}\rangle$,
 $|\frac{1}{2},\frac{1}{2}\rangle$, $|\frac{3}{2},\frac{1}{2}\rangle$, $|\frac{3}{2},\frac{-3}{2}\rangle$, has the form

\begin{equation}
 \label{eq:KL6-1}
 H_{\mb{\tiny KL}(6 \times 6)} = \left(
 \begin{array}{cccccc}
 P+Q & R & \sqrt{2} R & -\frac{S}{\sqrt{2}} & -S & 0 \\
 R^* & P-Q & \sqrt{2} Q & \sqrt{\frac{3}{2}} S^* & 0 & S \\
 \sqrt{2} R^* & \sqrt{2} Q & P+\Delta_{0}  & 0 & \sqrt{\frac{3}{2}} S^* & -\frac{S}{\sqrt{2}} \\
 -\frac{S^*}{\sqrt{2}} & \sqrt{\frac{3}{2}} S & 0 & P+\Delta_{0}  & -\sqrt{2} Q & -\sqrt{2} R \\
 -S^* & 0 & \sqrt{\frac{3}{2}} S & -\sqrt{2} Q & P-Q & R \\
 0 & S^* & -\frac{S^*}{\sqrt{2}} & -\sqrt{2} R^* & R^* & P+Q \\
\end{array}
\right),
\end{equation}
\noindent and after standard simplifications can be re-written as
\begin{equation}
 \label{eq:KL6-2}
 H_{\mb{\tiny KL}(6 \times 6)} = \left(
  \begin{array}{cccccc}
   \text{$H_{11}$} & \text{$H_{12}$} & \text{$H_{13}$} & \text{$H_{14}$} & \text{$H_{15}$} & 0 \\
   \text{$H_{12}$}^* & \text{$H_{22}$} & \text{$H_{23}$} & \text{$H_{24}$}^* & 0 & -\text{$H_{15}$} \\
   \text{$H_{13}$}^* & \text{$H_{23}$} & \text{$H_{33}$} & 0 & \text{$H_{24}$}^* & \text{$H_{14}$} \\
   \text{$H_{14}$}^* & \text{$H_{24}$} & 0 & \text{$H_{33}$} & -\text{$H_{23}$} & -\text{$H_{13}$} \\
   \text{$H_{15}$}^* & 0 & \text{$H_{24}$} & -\text{$H_{23}$} & \text{$H_{22}$} & \text{$H_{12}$} \\
   0 & -\text{$H_{15}$}^* & \text{$H_{14}$}^* & -\text{$H_{13}$}^* & \text{$H_{12}$}^* & \text{$H_{11}$} \\
 \end{array}
 \right);
\end{equation}
\noindent whose explicit matrix elements can be found in the \ref{app:KL6}.

\subsection{Quadratic Eigenvalue Problem}
 \label{subsec:QEP}
\subsubsection{Generalized Matrix Sturm-Liouville problem}

In the section \ref{sec:Intro}, various approximations where mentioned for describing the spin-charge carriers dynamics in strongly-coupled MMS. Whenever these theoretical models are invoked within the EFA for real MMS-layered heterostructures, more restrictions arise due the presence of topological requirements at the boundaries and at the interfaces. This type of problem is well-known as $N$-coupled components GSL matrix boundary-equation \cite{Pernas2015,malik,moli}, which for MMS heterostructures with translational symmetry in the $[x, y]$ plane of perfect interfaces, can be cast as follows

\begin{equation}
 \label{eq:SL}
 \frac{d}{dz}\left[\bn{\mathcal{B}}(z)\frac{d\mathbb{F}(z)}{dz}+\bn{\mathcal{P}}(z)\mathbb{F}(z)
 \right] + \bn{\mathcal{Y}}(z)\frac{d\mathbb{F}(z)}{dz} + \bn{\mathcal{W}}(z)\mathbb{F}(z) = \mathbb{O}_{N},
\end{equation}
\ni being $\bn{\mathcal{B}}(z), \bn{\mathcal{W}}(z)$ and $\bn{\mathcal{Y}}(z) = -\bn{\mathcal{P}}^{\dag}(z)$, $(N \times N)$ general-hermitian matrices \cite{Pernas2015,moli}. Hereinafter, $\mathbb{O}_{N}$ stands for the null $N$-order matrix. In equation (\ref{eq:SL}) the coordinate $z$ denotes the quantization direction, which is perpendicular to the $[x, y]$ plane and along with the momentum component $k_{z}$ is confined. Formally defined as the ``field" \cite{Perez2004}, $\mathbb{F}(z)$ in our problem represents an envelope $(N \times 1)$ spinor, whose $N$-component amplitudes will be derived from the QEP and can be treated independently as well as simultaneously in our approach. Some of the matrix-coefficients in (\ref{eq:SL}), depart from a strictly 1D $z$-dependent functions but also depend on the in-plane quasi-momentum $\vec{k}_{\mb{\tiny T}} = k_{x} \hat{e}_{x} + k_{y} \hat{e}_y$. For completeness worthwhile recalling that it is usual to call $\bn{\mathcal{B}}(z)$ as the mass-matrix, $\bn{\mathcal{P}}(z)$ as the dissipation-matrix (compiles interaction terms) and $\bn{\mathcal{W}}(z)$ as the
strain-matrix (includes the energy and the potential) \cite{Perez2004,goldstein}. A plain-wave solution of the form $\mathbb{F}(z) = \sum_{j=1}^{2N}\alpha_{j}e^{ik_{z_{j}}z}\varphi_{j}$ could be proposed in (\ref{eq:SL}), leading us to

\begin{equation}
 \label{eq:QEP}
 \mathbb{Q}(k_{z})\bn{\varphi} = \lbrace k_{z}^{2}\mathbb{M}+k_{z}\mathbb{C}+\mathbb{K}\rbrace \bn{\varphi}=\mathbb{O}_{N\times 1}
\end{equation}
\noindent which is a non-linear algebraic problem known as QEP \cite{tisseur} as it is straightforward that the $2^{nd}$-order $k_{z}$ \emph{eigen}-values are discrete quantities, while the $\kappa_{\mb{\tiny T}}$ components have continuous entries. Some properties of (\ref{eq:QEP}) will be outlined next.

\subsubsection{Outline of the QEP properties}
\label{subsec:QEP-Prop}

In the $2^{nd}$-order matrix polynomial Eq. (\ref{eq:QEP}), the matrixes $\mathbb{M},\mathbb{C}$ and $\mathbb{K}$ determine the $\mathbb{Q}(k_{z})$ spectrum \cite{tisseur}. In our case, they are assumed to be by-layer constants and hermitian to be consistent with similar assumption on the matrix-coefficients of the underlying GSL (\ref{eq:SL}). In particular, $\mathbb{M}$ is non-singular and regular, therefore the eigenvalues of $k_{z}$ are real or come in complex-conjugate pairs $(k_{z}, k_{z}^{*})$. These properties justify the \mb{\bm $\vec{k} \cdot \vec{p}$} approximation \cite{dsanchez,dimmock,coppol} in the EFA framework \cite{burt,foreman}. To find $k_{z}$ eigenvalues, we twice linearize (\ref{eq:QEP}), so the QEP acquires the form

\begin{equation}
 \label{eq:QEP-Lin1}
 \left[\begin{array}{cc}
        \mathbb{Q}(k_{z})&\mathbb{O}_{N}\\
        \mathbb{O}_{N}&\mathbb{I}_{N}
 \end{array}\right]=\mathbb{H}(k_{z})(\mathbb{A}-k_{z}\mathbb{B})\mathbb{G}(k_{z}),
\end{equation}
\noindent where $\mathbb{A}-k_{z}\mathbb{B}$ is $(2N\times 2N)$ lineal in $k_{z}$, and being $\mathbb{I}_{N}$ the $N$-order identity matrix. The $(2N\times 2N)$ matrixes $\mathbb{H}(k_{z})$ and $\mathbb{G}(k_{z})$ are related with the $\mathbb{Q}(k_{z})$ components   \cite{tisseur} and the $(\mathbb{A}-k_{z}\mathbb{B})$ eigenvalues are the same to those of $\mathbb{Q}(k_{z})$, provided that the substitution $\bn{\mu} = k_{z}\bn{\varphi}$ is a second-step linearization over $\mathbb{Q}(k_{z})$ \cite{tisseur}. Thus, (\ref{eq:QEP}) can be now re-written as

\begin{equation}
 \label{eq:QEP-Lin2}
 \left[\begin{array}{cc}
 \mathbb{O_{N}}&\mathbb{N}\\
 \mathbb{-K}&\mathbb{-C}
 \end{array}\right]\left[\begin{array}{c}
 \bn{\varphi}\\
 \bn{\mu}
 \end{array}\right]-k_{z}\left[\begin{array}{cc}
 \mathbb{N}&\mathbb{O_{N}}\\
 \mathbb{O_{N}}&\mathbb{M}
 \end{array}\right]\left[\begin{array}{c}
 \bn{\varphi}\\
 \bn{\mu}
 \end{array}\right]=\lbrace \mathbb{A}-k_{z}\mathbb{B}\rbrace \left[\begin{array}{c}
 \bn{\varphi}\\
 \bn{\mu}
 \end{array}\right]=\mathbb{O}_{2N\times 1},
\end{equation}
\ni which is understood as a generalized eigenvalue problem (GEP) \cite{lancaster,tisseur}, where $\mathbb{N}$ is any $(N \times N)$ non-singular matrix, often assumed after the $N$-order identity matrix $\mathbb{I}_{N}$. Although being double-sized respect to (\ref{eq:QEP}), the Eq.(\ref{eq:QEP-Lin2}) is more simple to solve because it can be established a canonical form, then one can obtain the eigenvalues analytically unlike the QEP. The last can be done by a generalized Schur decomposition (GSD) using the so called QZ algorithm \cite{qz} or by a simultaneous triangularization (STR) of the pencil  $(\mathbb{A},\mathbb{B})$, under certain  conditions \cite{mendoza,guillermo}, among other procedures \cite{Hammarling2013}. Indeed, the matrixes $\mathbb{A}$,$\mathbb{B}$ are simultaneous triangularizable if the matrixes $\mathbb{M}$ and $\mathbb{K}$ are invertible and:
\begin{equation}
 \label{cond1}
    \text{If} \;\;\mathbb{N}=\mathbb{M} \;\;\text{ and the matrix } \;\; \mathbb{K}\mathbb{N}\mathbb{K}^{-1}\mathbb{N}^{-1} \;\; \text{ are unipotent,} \footnote{\text{A matrix $\bn{U}$ is called unipotent if all eigenvalues are equal to $1$, or equivalently, a matrix $\bn{U}$ over a field is unipotent if and only if,} \\
    \text{its characteristic polynomial is $\left(u-1\right)^{n} = 0$, \emph{i.e.}, the matrix $\bn{U}-\mathbb{I}_{n}$ is nilpotent. A matrix $\bn{R}$ is called nilpotent if there exist an integer $n$,} \\
    \text{such that $\bn{R}^{n}= \mathbb{O}_{n}$}}
 \end{equation}
\begin{equation}
  \label{cond2}
   \text{or $\;\;$ If} \;\; \mathbb{N}=\mathbb{M} \;\; \text{ and } \;\; [\mathbb{N},\mathbb{K}]=\mathbb{N}\mathbb{K}-\mathbb{K}\mathbb{N} = \mathbb{O}_{N}.
\end{equation}

As mentioned above, this approach had been used successfully for KL models with $N = 2,4$ \cite{mendoza,job}, though without the interaction effects of the \textit{sh}. We claim that sub-bandmixing effects strongly modulated by such spin-dependent quasi-particles, could be remarkable for the development of technological nano-spintronics devices \cite{xiangyu,winklerspin,doty,Park2014,Kumar2017}. For this reason, a more detailed analysis of the sub-bandmixing consequences together with the interplay of \textit{hh-sh} and \textit{lh-sh}, in the extended KL model will be shown soon after.  This later attempt we perform both, \emph{via} the GSD as well as by validating the conditions (\ref{cond1}) and (\ref{cond2}) provided the STR reliability for the derived $(12 \times 12)$ GEP.

\subsubsection{QEP for the ($6 \times 6$) KL model}

Once we have defined $H_{\mb{\tiny KL}(6 \times 6)}$ from (\ref{eq:KL6-2}), next we need to link the matrixes $\mathbb{M},\mathbb{C}$ and $\mathbb{K}$ with the GSL Eq. (\ref{eq:SL}). By doing so we get: $\mathbb{M} = -\bn{\mathcal{B}}, \mathbb{C} = 2i\bn{\mathcal{P}}$ and $\mathbb{K} = \bn{\mathcal{W}}$ \cite{Perez2004} and therefore

\begin{flalign}
 \label{eq:C-QEP6}
 \mathbb{C}=\left(
  \begin{array}{cccccc}
   0 & 0 & 0 & \mathcal{H}_{14} & \mathcal{H}_{15} & 0 \\
   0 & 0 & 0 & \mathcal{H}_{24}{}^* & 0 & -\mathcal{H}_{15} \\
   0 & 0 & 0 & 0 & \mathcal{H}_{24}{}^* & \mathcal{H}_{14} \\
   \mathcal{H}_{14}{}^* & \mathcal{H}_{24} & 0 & 0 & 0 & 0 \\
   \mathcal{H}_{15}{}^* & 0 & \mathcal{H}_{24} & 0 & 0 & 0 \\
   0 & -\mathcal{H}_{15}{}^* & \mathcal{H}_{14}{}^* & 0 & 0 & 0 \\
  \end{array}
 \right),
\end{flalign}

\begin{flalign}
 \label{eq:M-QEP6}
  \mathbb{M}=\left(
   \begin{array}{cccccc}
    m_1 & 0 & 0 & 0 & 0 & 0 \\
    0 & m_2 & m_s & 0 & 0 & 0 \\
    0 & m_s & m_3 & 0 & 0 & 0 \\
    0 & 0 & 0 & m_4 & -m_s & 0 \\
    0 & 0 & 0 & -m_s & m_5 & 0 \\
    0 & 0 & 0 & 0 & 0 & m_6 \\
   \end{array}
  \right),
\end{flalign}

\begin{flalign}
 \label{eq:K-QEP6}
  \mathbb{K}=\left(
   \begin{array}{cccccc}
     a_1 & H_{12} & H_{13} & 0 & 0 & 0 \\
     H_{12}{}^* & a_2 & a_s & 0 & 0 & 0 \\
     H_{13}{}^* & a_s & \mathcal{H}_{33} & 0 & 0 & 0 \\
     0 & 0 & 0 & \mathcal{H}_{33} & -a_s & -H_{13} \\
     0 & 0 & 0 & -a_s & a_2 & H_{12} \\
     0 & 0 & 0 & -H_{13}{}^* & H_{12}{}^* & a_1 \\
\end{array}
\right),
\end{flalign}
\noindent with matrix elements explicitly presented in the \ref{app:QEP6}. From expressions (\ref{eq:C-QEP6})-(\ref{eq:K-QEP6}) is straightforward that the obtained QEP is regular and non-singular, therefore twelve finite-real or six complex-conjugate pairs of eigenvalues are expected. By comparing with the case of the KL model with $N=4$ the matrixes $\mathbb{M},\mathbb{C}$ and $\mathbb{K}$ present some differences due to the inclusion of the spin orbit (SO) band and the interaction between \textit{lh}, \textit{hh} and \textit{sh}. The most remarkable though, is the appearance of some mixing-free off-diagonal terms $m_{s}$ in the mass-matrix $\mathbb{M}$. Therefore, the \textit{hh-sh} and \textit{lh-sh} interactions \cite{ekbote} becomes independent of any bandmixing regime. This interplay will be described soon after in Sec. \ref{sec:Sim} by testing the influence of $m_{s}$ on QEP spectral distribution. Assuming the axial approximation \cite{vurgaftman}, the effective masses were taken after: $m_{hh}^{*} = m_{0}/(\gamma_{1}-2\gamma_{2}), \;m_{lh}^{*} = m_{0}/(\gamma_{1}+2\gamma_{2})\,$ and $\,m_{sh}^{*} = m_{0}/(\gamma_{1}-\gamma_{so})$. Here $m_{0}$ stands for the bare electron mass, and $\gamma_{so} = E_{P}\Delta_{so}/(3E_{g}(E_{g} + \Delta_{so}))$, being $E_{P}=2m_{0}P^{2}/\hbar^{2}$, where $P = -i\hbar \bra{S}p_{x}\ket{X}/m_{0}$ represents the momentum matrix element between the \textit{s}-like conduction bands and \textit{p}-like valence bands \cite{vurgaftman} (for $GaAs(AlAs)$ have been taken $E_{p} = 28.8(21.1)$ eV, respectively). Worthwhile to remark that besides the standardized effective-mass dependence of the semi-empirical L\"uttinger parameters, the SO subband-gap $\Delta_{0}$ will have an important role on determining their values.

From (\ref{eq:QEP-Lin2}) by taking $\mathbb{N}$ with non-zero diagonal elements we have

\begin{equation}
 \label{eq:A-GEP12}
  \mathbb{A} = \left[\begin{array}{cccccccccccc}
   0 & 0 & 0 & 0 & 0 & 0 & n_{1} & 0  & 0 & 0 & 0 & 0\\
   0 & 0 & 0 & 0 & 0 & 0 & 0 & n_{2} & 0 & 0 & 0 & 0\\
   0 & 0 & 0 & 0 & 0 & 0 & 0 & 0 & n_{3} & 0 & 0 & 0\\
   0 & 0 & 0 & 0 & 0 & 0 & 0 & 0 & 0 & n_{4} & 0 & 0\\
   0 & 0 & 0 & 0 & 0 & 0 & 0 & 0 & 0 & 0 & n_{5} & 0\\
   0 & 0 & 0 & 0 & 0 & 0 & 0 & 0 & 0 & 0 & 0 & n_{6}\\
   -a_{1} & -H_{12} & -H_{13} & 0 & 0 & 0 & 0 & 0 & 0 & -\mathcal{H}_{14} & -\mathcal{H}_{15} & 0\\
   -H_{12}^{*} & -a_{2} & -a_{s} & 0 & 0 & 0 & 0 & 0 & 0 & -\mathcal{H}_{24}^{*} & 0 & \mathcal{H}_{15}\\
   -H_{13}^{*} & -a_{s} & -\mathcal{H}_{33} & 0 & 0 & 0 & 0 & 0 & 0 & 0 & -\mathcal{H}_{24}^{*} & -\mathcal{H}_{14}\\
   0 & 0 & 0 & -\mathcal{H}_{33} & a_{s} & H_{13} & -\mathcal{H}_{14}^{*} & -\mathcal{H}_{24} & 0 & 0 & 0 & 0\\
   0 & 0 & 0 & a_{s} & -a_{2} & -H_{12} & -\mathcal{H}_{15}^{*} & 0 & -\mathcal{H}_{24} & 0 & 0 & 0\\
   0 & 0 & 0 & H_{13}^{*} & -H_{12}^{*} & -a_{1} & 0 & \mathcal{H}_{15}^{*} & -\mathcal{H}_{14}^{*} & 0 & 0 & 0
  \end{array}\right],
\end{equation}
\begin{equation}
 \label{eq:B-GEP12}
  \mathbb{B} = \left[\begin{array}{cccccccccccc}
   n_{1} & 0 & 0 & 0 & 0 & 0 & 0 & 0 & 0 & 0 & 0 & 0\\
   0 & n_{2} & 0 & 0 & 0 & 0 & 0 & 0 & 0 & 0 & 0 & 0\\
   0 & 0 & n_{3} & 0 & 0 & 0 & 0 & 0 & 0 & 0 & 0 & 0\\
   0 & 0 & 0 & n_{4} & 0 & 0 & 0 & 0 & 0 & 0 & 0 & 0\\
   0 & 0 & 0 & 0 & n_{5} & 0 & 0 & 0 & 0 & 0 & 0 & 0\\
   0 & 0 & 0 & 0 & 0 & n_{6} & 0 & 0 & 0 & 0 & 0 & 0\\
   0 & 0 & 0 & 0 & 0 & 0 & m_{1} & 0 & 0 & 0 & 0 & 0\\
   0 & 0 & 0 & 0 & 0 & 0 & 0 & m_{2} & m_{s} & 0 & 0 & 0\\
   0 & 0 & 0 & 0 & 0 & 0 & 0 & m_{s} & m_{3} & 0 & 0 & 0\\
   0 & 0 & 0 & 0 & 0 & 0 & 0 & 0 & 0 & m_{4} & -m_{s} & 0\\
   0 & 0 & 0 & 0 & 0 & 0 & 0 & 0 & 0 & -m_{s} & m_{5} & 0\\
   0 & 0 & 0 & 0 & 0 & 0 & 0 & 0 & 0 & 0 & 0 & m_{6}\\
  \end{array}\right]
\end{equation}

From expressions (\ref{eq:A-GEP12}) and (\ref{eq:B-GEP12}) we can conclude that the $det(\mathbb{A})\neq 0$ and $det(\mathbb{B})\neq 0$ conserving then all required properties of the QEP. To obtain the corresponding $(12 \times 12)$-GEP eigenvalues through a GSD, the following algorithm (\ref{alg:L1}) was performed

\begin{algorithm}
 \caption{Eigenvalue obtention of the QEP for the KL ($N\times N$)}
 \label{alg:L1}
 \begin{algorithmic}[L1(1)]
  \State Start \Comment{\textit{\small{Define parameters ($k_{x}$, $k_{y}$, $k_{T}$, $\gamma_{i}$, $V$, $E$ and $\Delta_{0}$)}}}
  \State Obtain matrixes $\mathbb{M}$, $\mathbb{C}$, $\mathbb{K}$ and $\mathbb{N}$
  \Procedure{Eig\_QEP}{$\mathbb{M}$,$\mathbb{C}$,$\mathbb{K}$,$\mathbb{N}$}
  \State Define matrices $\mathbb{A}$ and $\mathbb{B}$ with Eq. (\ref{eq:A-GEP12})-(\ref{eq:B-GEP12})
  \If{$det(\mathbb{A})\neq 0$ \& $det(\mathbb{B})\neq0$}
  \State Solve Eq. (\ref{eq:QEP-Lin2})
  \State Eigenvalue obtention by GSD \cite{qz}
  \State $\lambda_{i}={T_{A}}_{ii}/{T_{B}}_{ii}$
  \EndIf \Comment{\textit{For the GSD we must assure that $\mathbb{A}=\mathbb{Q}T_{A}\mathbb{Z}^{\dagger}$ and $\mathbb{B}=\mathbb{Q}T_{B}\mathbb{Z}^{\dagger}$, where $\mathbb{Q}$ and $\mathbb{Z}$ are unitary matrices and $T_{A}$ and $T_{B}$ are triangular matrices, then the eigenvalues are given by  $\lambda_{i}={T_{A}}_{ii}/{T_{B}}_{ii}$}}
  \State return $\lambda_{i}$
  \EndProcedure
  \State End
 \end{algorithmic}
\end{algorithm}

\subsection{Validation of the sufficient conditions for a STR}

As we have mentioned in subsection \ref{subsec:QEP-Prop}, the STR can be implemented for the obtention of the respective eigenvalues, if certain imposed conditions \cite{mendoza,guillermo} are accomplished. Thus, the rules (\ref{cond1}) or (\ref{cond2}), must be verified to determine if the STR applies for the extended $(6\times6)$ KL model. For the case of the condition (\ref{cond1}) we take the expressions (\ref{eq:M-QEP6})-(\ref{eq:K-QEP6}) and evaluate $\mathbb{K}\mathbb{N}\mathbb{K}^{-1}\mathbb{N}^{-1}$. Provided $\mathbb{N} = \mathbb{M}$, the characteristic polynomial is of the form $(1-\lambda)^{6}=q_{6}\lambda^{6}+q_{5}\lambda^{5}+\dots+q_{1}\lambda+q_{0}$,  whose eigenvalues are function of $\mathbb{K}$ and $\mathbb{M}$ matrix elements. Besides, it is mandatory that $\lambda_{i} = 1$. We have tested (\ref{cond1}) for $GaAs$ and $AlAs$, obtaining the results shown in Fig.\ref{3d_sms}. As can be observed, the requirement (\ref{cond1}) is not satisfied, for both semiconducting alloys. For the $GaAs$ --for example--, we have obtained:  $Log|Re(\lambda)|=[-0.6942;-0.6935]\Rightarrow|Re(\lambda)|\neq 1$ and $Log|Im(\lambda)|=[9.095\times 10^{-3};-9.065\times 10^{-3}]\Rightarrow|Im(\lambda)|\neq 0$. Let us now turn to the condition (\ref{cond2}), whose commutator reads

{\small
\begin{equation}
 \label{for: conmu}
 [\mathbb{N},\mathbb{K}]=
 \left[
\begin{array}{cccccc}
 0 & H_{12}(n_{1}-n_{2}) & H_{13} (n_{1}-n_{3}) & 0 & 0 & 0 \\
 -H_{12}^{*} (n_{1}-n_{2}) & 0 & a_{s} (n_{2}-n_{3}) & 0 & 0 & 0 \\
 -H_{13}^{*} (n_{1}-n_{3}) & a_{s} (n_{3}-n_{2}) & 0 & 0 & 0 & 0 \\
 0 & 0 & 0 & 0 & a_{s}(n_{5}-n_{4}) & -H_{13} (n_{4}-n_{6}) \\
 0 & 0 & 0 & a_{s} (n_{4}-n_{5}) & 0 & H_{12} (n_{5}-n_{6}) \\
\end{array}
\right].
\end{equation}
}

By testing again for $GaAs$ and $AlAs$, we show in Fig.\ref{con_sms} the spectral norm of (\ref{for: conmu}) with $\mathbb{N}=\mathbb{M}$. As can be seen, the condition (\ref{cond2}) is not satisfied neither, because the values of the spectral norm must be approximately zero. Importantly, after some algebra around the commutator (\ref{for: conmu}) keeping  $\mathbb{N}=\mathbb{M}$, we found that for the fulfillment of the sufficient conditions (\ref{cond1}) and (\ref{cond2}), the off-diagonal term $m_{s}$ in (\ref{eq:M-QEP6}) have to be zeroed. We then impose that: $\,n_{1} = n_{2}$, $n_{3} = n_{2}$, $n_{6} = n_{1}$, $n_{5} = n_{2}$ and $n_{3} = n_{4}$. Then we get: $\,m_{1} = m_{6} = m_{2} = m_{5}$ and $m_{3} = m_{4} = m_{2}$. Thereby, the above defined effective masses should be reformulated as $m_{hh}^{*}\approx m_{lh}^{*}\;$ and $\;m_{sh}^{*}\approx m_1H_{13}-m_{s}H_{12}\approx m_{lh}^{*}$. Now, by re-evaluating the sufficient conditions (\ref{cond1}) and (\ref{cond2}) for the STR, but with  $m_{s} \cong 0$ we obtain for (\ref{cond1}) the results displayed in Fig.\ref{3d_ms}, where it can be observed that for both the $GaAs$ and $AlAs$ the unipotent restriction is fulfilled. In the case of the GaAs, we found $Log|Re(\lambda)|=[-0.2\times10^{-8};-1.4\times10^{-8}]\Rightarrow|Re(\lambda)|\approx 1$ and $Log|Im(\lambda)|=[-4;-9]\Rightarrow|Im(\lambda)|\approx 0$. This result implies that by varying the in-plane parameters $k_{x}$ and $k_{y}$ the eigenvalues remain intact. Therefore, the pencil ($\mathbb{A}$,$\mathbb{B}$) is subjected to STR. For the requirement (\ref{cond2}), we have presented in Fig.\ref{con_ms}, that  the commutator (\ref{for: conmu}) is approximately zero, so this condition is also fulfilled for $GaAs$ and $AlAs$. However, we emphasize that only when $m_{s}$ approaches zero, both sufficient conditions fulfill, which implies a very restrictive relation for the effective masses, \textit{i.e.} $m_{hh} \approx m_{lh}^{*} \approx m_{sh}^{*}$. This similarity is rather far from real materials with wide technological applications. For that reason in the next section, we will consider the GSD method to obtain the eigenvalues.

\section{Numerical Simulations and Discussions}
\label{sec:Sim}

The results presented below, where obtained by using the algorithm (\ref{alg:L1}) for the quotation of the GEP eigenvalues. Next, for the \textit{root-locus} plots we use the algorithm (\ref{alg:L2}) taking the scattering potential as $V = 0.498$ eV for $AlAs$ and correspondingly $V = 0$ eV for $GaAs$. The band mixing parameter $\kappa_{\mb{\tiny T}} \in  [10^{-6},10^{-1}]$\AA$^{-1}$, where we have assumed the intervals $[10^{-6},10^{-4}]$\AA$^{-1}$ as low-mixing regime; $[10^{-4},10^{-3}]$\AA$^{-1}$ as moderate-mixing regime and $[10^{-3},10^{-1}]$\AA$^{-1}$ as high-mixing regime. We exercise the \textit{root-locus-like} technique by showing the influence of the term $m_{s}$ on the $(12 \times 12)$ GEP (\ref{eq:QEP-Lin2}) spectral distribution as $\kappa_{\mb{\tiny T}}$ increases. Finally, we have made a comparison between $hh-sh$ and $lh-sh$ cases.

\subsection{QEP Spectral Distribution}

For a better interpretation of the charge-carriers eigenvalues, we retrieve the \emph{root-locus-like} procedure, provided its success to directly analyze specific physical phenomena involving uncoupled and/or coupled modes of MMS\cite{mendoza,job}. On the ground of the classical control theory, we underline the remarkable graphical resemblance from Evans' approach \cite{Evans48} for the study of dynamical systems, widely known as \emph{Root-Locus}. The \emph{root-locus-like} technique allows to graph the eigenvalues evolution of the characteristic polynomial in the complex plane and thus we are able to follow the system stability criteria. For our modeling --\textit{via} the algorithm (\ref{alg:L2})--, we will plot the GEP (\ref{eq:QEP-Lin2}) eigenvalues to characterize the propagating and evanescent modes of the charge carriers (\textit{lh}, \textit{hh} and \textit{sh}), mainly for bulk binary-compound semiconductors, such as $GaAs$ and $AlAs$. Nonetheless, we guess that our method is suitable for other specialized $III-V$ and $II-VI$ semiconducting alloys with minor changes, if any.

\begin{figure} [tbp]
 \centering
\subfloat[$H_{KL (4\times 4)} $]{
\label{fig:Ga4x4rl}
   \includegraphics[width=0.40\textwidth]{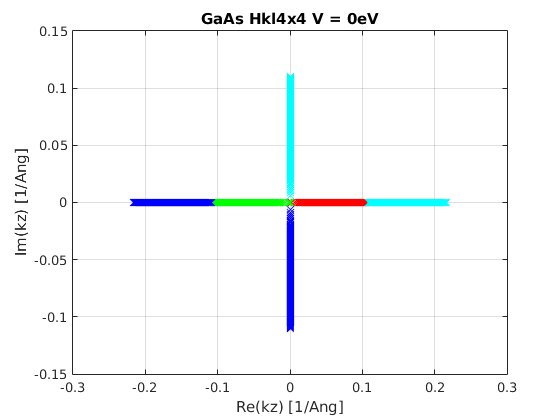}}
   \subfloat[$H_{KL (4\times 4)}$]{
\label{fig:al4x4rl}
   \includegraphics[width=0.40\textwidth]{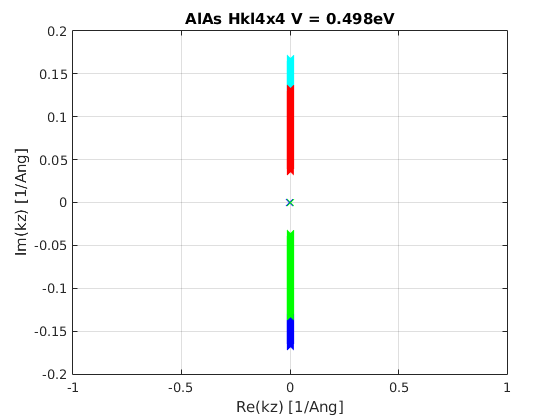}}\\
\subfloat[$H_{KL (6\times 6)},m_{s}=0$]{
\label{fig:Ga6x6rl}
   \includegraphics[width=0.40\textwidth]{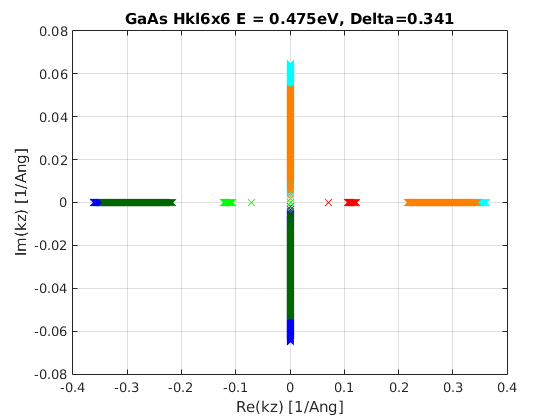}}
\subfloat[$H_{KL (6\times 6)}$]{
\label{fig:Ga6x6msrl}
   \includegraphics[width=0.40\textwidth]{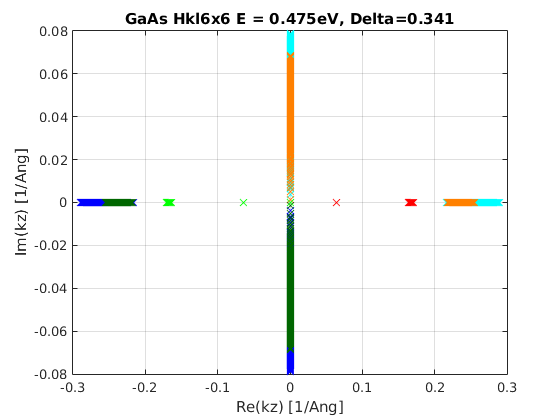}}\\
   \subfloat[$H_{KL (6\times 6)},m_{s}=0$]{
\label{fig:al6x6rl}
   \includegraphics[width=0.40\textwidth]{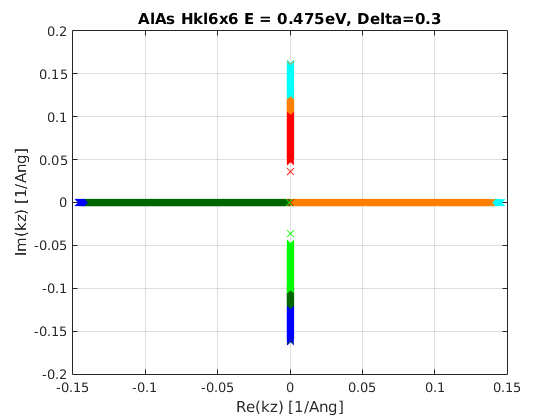}}
\subfloat[$H_{KL (6\times 6)}$]{
\label{fig:al6x6msrl}
   \includegraphics[width=0.40\textwidth]{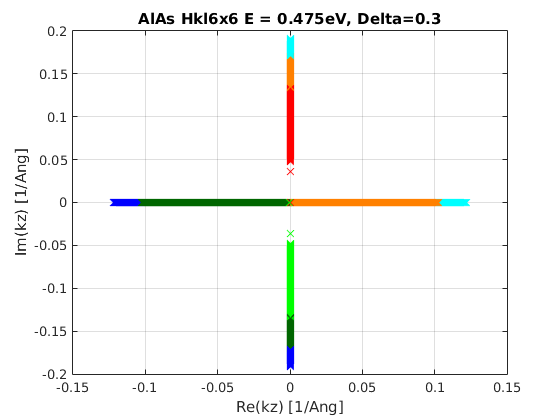}}
\caption[graf]
   {\label{fig:GaAsrl} (Color online) Plot of the \textit{root-locus-like} procedure for $k_{z}$ eigenvalues of [$hh$(\textcolor{bluel}{x}-\textcolor{blues}{x}), $lh$(\textcolor{red}{x}-\textcolor{green}{x}), $sh$(\textcolor{orange}{x}-\textcolor{gdark}{x})], with $\kappa_{\mb{\tiny T}}$. We have considered $GaAs$($AlAs$) with $V = 0 (0.498)$ eV, respectively. Have been taken $\kappa_{\mb{\tiny T}}$  $\in [10^{-6},10^{-1}]$ \AA$^{-1}$ and the incident energy $E = 0.475$ eV.}
\end{figure}

In Fig.\ref{fig:GaAsrl} we have used the \textit{root-locus-like} technique in $GaAs$ and $AlAs$, to graph the evolution of the GEP (\ref{eq:QEP-Lin2}) eigenvalues as $\kappa_{\mb{\tiny T}}$ increases. Panels (\ref{fig:Ga4x4rl}) and (\ref{fig:al4x4rl}) revisit the known case for KL with $N=4$, to show how the eigenvalues of \textit{hh} (\textcolor{bluel}{x}-\textcolor{blues}{x}) and \textit{lh} (\textcolor{red}{x}-\textcolor{green}{x}) evolve from real to imaginary axis; \textit{i.e.} from propagating  to evanescent modes, respectively,  as $\kappa_{\mb{\tiny T}}$ grows. A phenomenology of this sort have been described before, whose direct consequence is the interplay of the effective scattering potential that is ``felt" by the charge-carrier as it travels throughout the heterostructure trespassing allowed QW-acting or forbidden QB-acting layers, respectively, whenever the band mixing parameter changes \cite{Diago2006,mendoza,job}. However, by taking into account the SO interaction in the extended KL model with $N = 6$, we have found a more cumbersome behavior. Panel (\ref{fig:Ga6x6rl}) charts the influence of the $sh$ (\textcolor{orange}{x}-\textcolor{gdark}{x}) states, on the behavior of $hh$ and $lh$ ones, disregarding the term $m_{s}$. Indeed, by comparing panel (\ref{fig:Ga6x6rl}) with panel (\ref{fig:Ga6x6msrl}) --taking into account $m_{s}$--, one observes an increased number of $sh$ states having a lower value $k_{z}$ for $\Re(k_{z}) \approx \left|\left[0.21,0.25\right]\right|$ \AA$^{-1}$, while for $\Im(k_{z}) \approx \left|\left[0.01,0.07\right]\right|$ \AA$^{-1}$ it becomes bigger. Therefore, since the $sh$ states evolve toward the imaginary axis at certain $\kappa_{\mb{\tiny T}}$ value, these quasi-particles ``feel'' a metamorphosis of the scattering potential from a QW-acting type into an effective QB-acting layer, in short, they behave as $hh$. By observing the panel (\ref{fig:Ga6x6rl}), we have detected more $sh$ states at a lower mixing parameter than those we have found from panel (\ref{fig:Ga6x6msrl}), because they change to imaginary values at $\kappa_{\mb{\tiny T}} \approx 0.09$ \AA$^{-1}$ [see panel (\ref{fig:Ga6x6rl})] and $\kappa_{\mb{\tiny T}} \approx 0.11$ \AA$^{-1}$ [see panel (\ref{fig:Ga6x6msrl})]. Interestingly, for $\kappa_{\mb{\tiny T}} \approx 9.6 \times 10^{-2}$ \AA$^{-1}$ the $k_{z,sh}$ eigenvalues take imaginary entries, which is approximately the same $\kappa_{\mb{\tiny T}}$ for $k_{z,lh}$ within the $(4\times4)$ KL model \cite{mendoza}.

\begin{algorithm}
\caption{\textit{root-locus-like} technique}
 \label{alg:L2}
 \begin{algorithmic}[L1(2)]
 \State Start \Comment{\textit{\small{We take into account algorithm (\ref{alg:L1})}}}
\Procedure{RL}{\textit{Known parameters}} \Comment{\textit{Defined in algorithm (\ref{alg:L1})}}
\State $k_{x}=k_{y}$\Comment{\textit{As a particular case}}
\State $\kappa_{\mb{\tiny T}}\gets \sqrt{k_{x}^{2}+k_{y}^{2}}$
\For{$\kappa_{\mb{\tiny T}}=0,0.1$}\Comment{\textit{\small{Take the right fixed step for the \textit{for} loop}}}
\State $\lambda_{i} \gets$ EIG\_QEP \Comment{\textit{ Algorithm \ref{alg:L1} procedure is used}}
\State Take the Real and Imaginary part of $\lambda_{i}$ ($Re(\lambda_{i}$) and $Im(\lambda_{i})$)
\State Separate the eigenvalues of \textit{hh}, \textit{lh} and $sh$  ($hh \leq sh \leq lh$)
\State Plot $Re(\lambda_{i})$ Vs $Im(\lambda_{i})$\Comment{\textit{\small{Differing each quasi-particle}}}
\State Plot Eigenvalues Vs $\kappa_{\mb{\tiny T}}$\Comment{\textit{QW-QB Profiles}}
\EndFor
\EndProcedure
\State End \hspace{18mm} The acronym QW(QB), hereinafter stands for quantum well(barrier).
\end{algorithmic}
\end{algorithm}

This produces a change from propagating modes to evanescent ones for the $sh$ and $hh$ states. On the contrary the $lh$ modes remain real and thus preserve their propagating character. Panel (\ref{fig:Ga6x6msrl}) shows the eigenvalue evolution for the $hh$, $lh$ and $sh$ considering the term $m_{s}$ for $GaAs$. Notice the remarkable different dynamic in comparison with that of the ($4 \times 4$) KL model [see panels (\ref{fig:Ga4x4rl}) and (\ref{fig:al4x4rl})]. We guess that the $sh$-related \textit{off}-diagonal elements in the GEP mass-matrix (\ref{eq:B-GEP12}), lead the $hh-sh-lh$ spectral distribution to change in the interval $\left|\left[0.01,0.1\right]\right|$ {\AA}$^{-1}$ --mostly seen for $k_{z,sh}$--, due to an additional increment of $\kappa_{\mb{\tiny T}}$. Considering $k_{z,hh}$ eigenvalues, for example, we have found they are confined to the interval $\left|[0.21; 0.3]\right|$ {\AA}$^{-1}$. In this sense, we had detected a moderate shift of the $k_{z,lh}$ eigenvalues towards those of $sh$. Besides, the effective masses will be also influenced by $m_{s}$, creating a greater similarity between $m_{hh,lh,sh}^{*}$ as a function of the Lüttinger parameter $\gamma_{2}$. Finally, worthwhile recalling the strong influence of $\Delta_{0}$ on the quotation of $sh$ eigenvalues. Panel (\ref{fig:al6x6rl}) charts the evolution of the $hh$  (\textcolor{
bluel}{x}-\textcolor{blues}{x}), \textit{lh} (\textcolor{red}{x}-\textcolor{green}{x}) and \textit{sh} (\textcolor{orange}{x}-\textcolor{gdark}{x}) for $AlAs$, disregarding $m_s$. Here, we can see that as for $GaAs$, [see panel (\ref{fig:al6x6msrl})] there is a large number of \textit{sh} states at low $\Re(k_{z}) \approx \left|\left[0 - 0.11\right]\right|$ \AA$^{-1}$ and at high $\Im(k_{z}) \left|\left[0.14 - 0.18\right]\right|$ \AA$^{-1}$. Thus for $AlAs$ some eigenvalues starting in the imaginary axis (evanescent modes) move later towards the real one (propagating modes), meanwhile other eigenvalues beginning in the neighborhood of $\kappa_{\mb{\tiny T}} \approx 0.015$ \AA$^{-1}$ take real values solely. It is worth noticing that the last behavior we have observed only for $sh$ and $hh$, when their eigenvalues approach or get away by varying $m_{s}$. Thereby it may be a relation between them. Panel (\ref{fig:al6x6msrl}) shows the same for $AlAs$, taking into account the SO band and the term  $m_s$. We found an analogous feature to that discussed above [see panel (\ref{fig:al6x6rl})], \emph{i.e.}, the domain of the $k_{z}$-eigenvalues is bigger for the imaginary part than that for the real one. We have also noticed, that the $\Im(k_{z,lh})$ eigenvalues vary more for $k_{z} \in \left|\left[0.05 - 0.14\right]\right|$ \AA$^{-1}$ in comparison with those of $sh$ and $hh$, whenever $k_{z} \in \left|\left[0.14 - 0.17\right]\right|$ \AA$^{-1}$ and $k_{z} \in \left|\left[0.17 - 0.19\right]\right|$ \AA$^{-1}$ respectively. In short words, we consider the $lh$ evanescent modes become more $\kappa_{\mb{\tiny T}}$-dependent, since its eigenvalues evolve from pure imaginary towards real valued, changing as well the character of the involved scattering modes.

\begin{figure}[tbp]
 \begin{center}
  \subfloat[QW(QB)-acting profiles at top(bottom) panel.]{
  \label{fig:Ga6x6mssep}
  \includegraphics[width=0.40\textwidth]{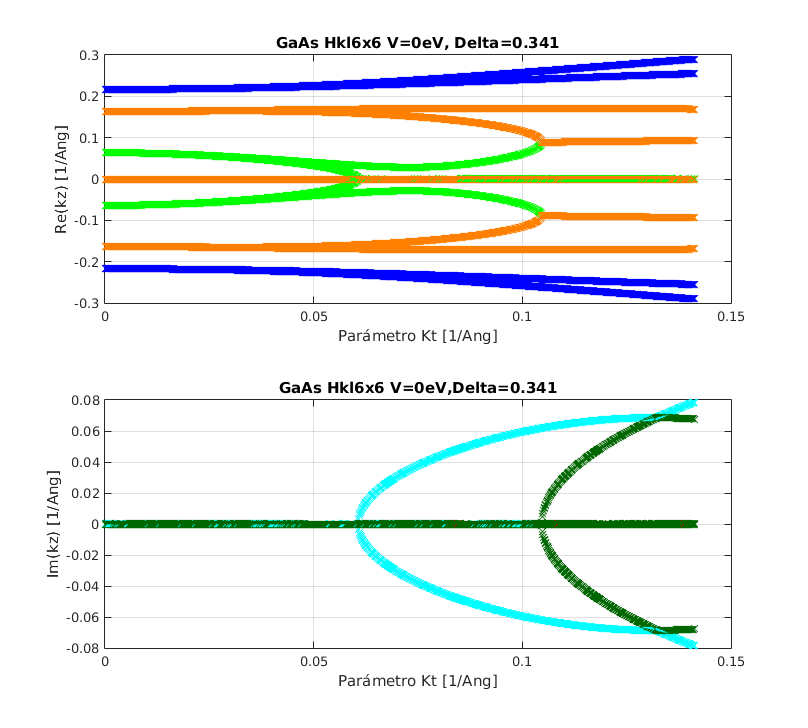}}
    \subfloat[Superposition of the QW(QB)-acting profiles.]{
  \label{fig:Ga6x6mspb}
  \includegraphics[width=0.40\textwidth]{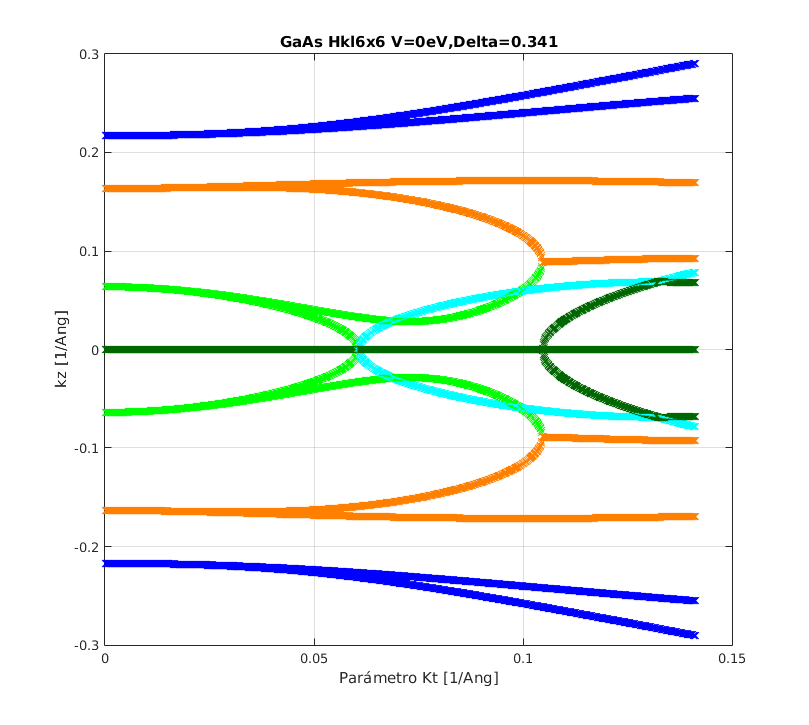}}\\
    \subfloat[QW(QB)-acting profiles at top(bottom) panel.]{
  \label{fig:Ga6x6sep}
  \includegraphics[width=0.40\textwidth]{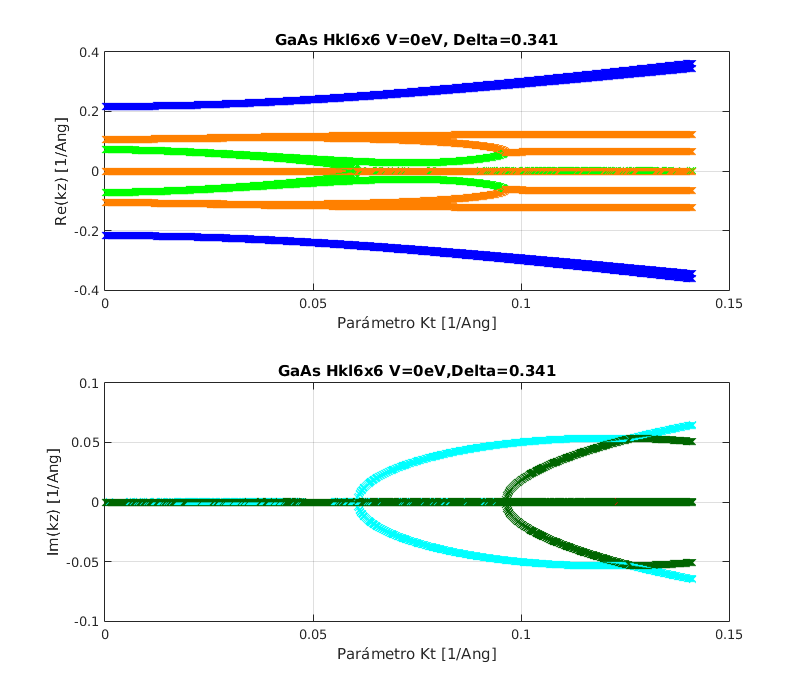}}
    \subfloat[Superposition of the QW(QB)-acting profiles.]{
  \label{fig:Ga6x6pb}
  \includegraphics[width=0.40\textwidth]{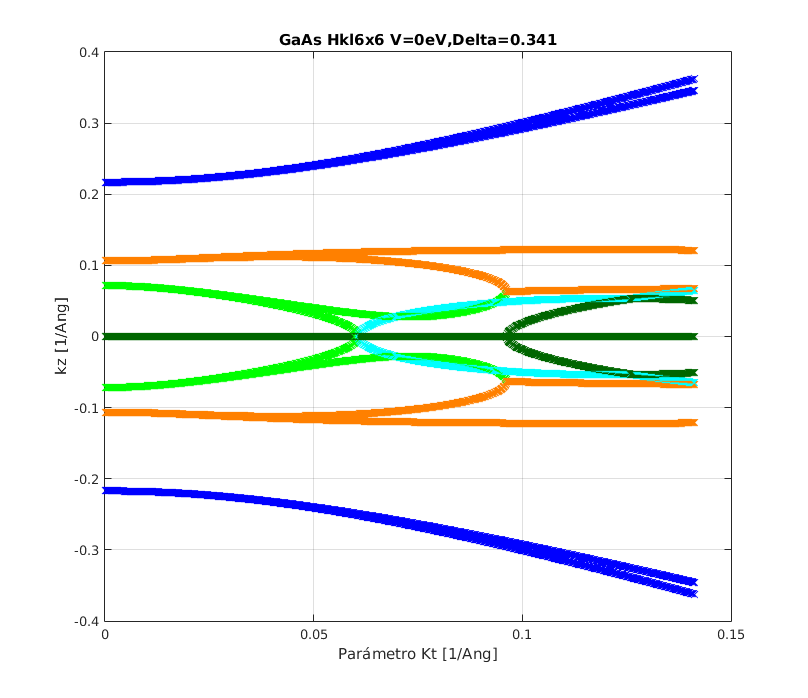}}\\
  \subfloat[QW(QB)-acting profiles at top(bottom) panel.]{
  \label{fig:Ga4x4sep}
  \includegraphics[width=0.40\textwidth]{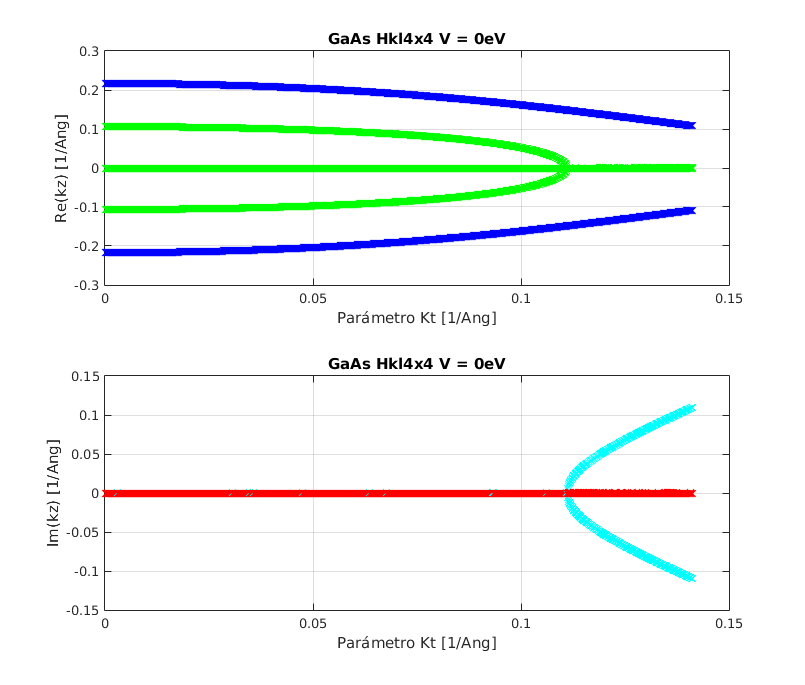}}
  \subfloat[Superposition of the QW(QB)-acting profiles.]{
  \label{fig:Ga4x4pb}
  \includegraphics[width=0.40\textwidth]{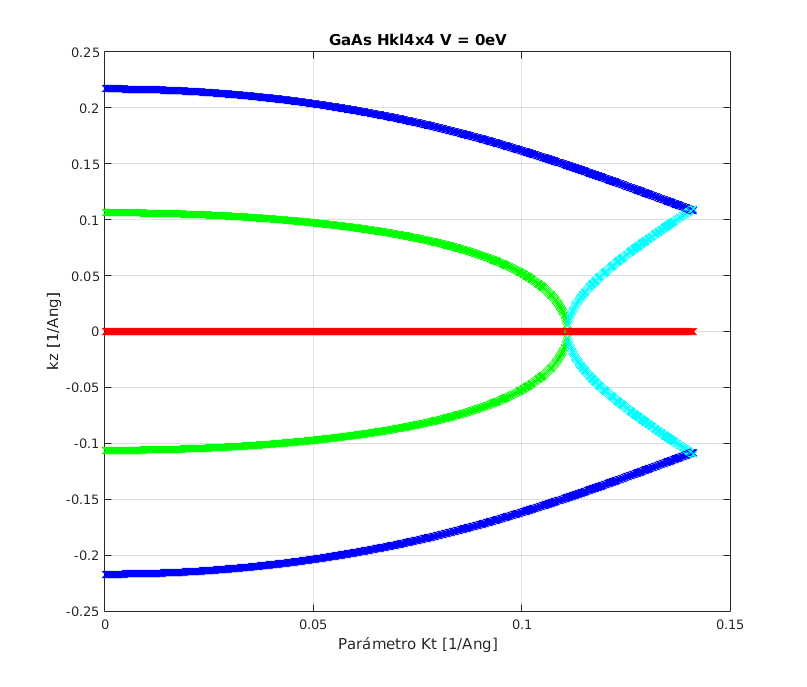}}
 \end{center}
\caption[6vs4al]{\label{GaAspb} (Color online) Plot of the \textit{root-locus-like} procedure for $k_{z}$ eigenvalues of [$hh$(\textcolor{bluel}{x}-\textcolor{blues}{x}), $lh$(\textcolor{red}{x}-\textcolor{green}{x}), $sh$(\textcolor{orange}{x}-\textcolor{gdark}{x})], as a function of $\kappa_{\mb{\tiny T}}$. We have considered $GaAs$ with $V = 0$ eV, respectively. For panels III-IV we have taken $m_{s} = 0$. We assumed $\kappa_{\mb{\tiny T}}$  $\in [10^{-6},10^{-1}]$ \AA$^{-1}$.}
\end{figure}

\subsection{QEP spectral distribution profiles}
\label{subsec:profiles}

Previous theoretical studies in the framework of the ($6 \times 6$) KL model, have disregarded \cite{ekbote} or even zeroed the \textit{sh}-related \textit{off}-diagonal elements in the QEP mass-matrix \cite{harrison2005}. It is noteworthy that in describing the ($6 \times 6$) KL approach some authors have derived an analogous non-linear QEP, however the \textit{sh}-related \textit{off}-diagonal elements contribution is not considered. Instead, they just mention the problem and propose a clue to solve it \cite{harrison2005}, which is by the way, different to our method. Next, we spread some light to this intricate issue by providing several evidences confirming the influence of the terms $m_{s}$ from (\ref{eq:B-GEP12}) on the $hh-lh-sh$ spectral distribution profiles and properties.

Fig.\ref{GaAspb} shows for $GaAs$, the explicit evolution of the eigenvalues from QEP (\ref{eq:QEP}) as a function of $\kappa_{\mb{\tiny T}}$ taking into account the terms $m_{s}$. We pursue a better observation  of the way the propagating and evanescent modes of the quasi-particles are modified through the scattering potential as the mixing between subbands increases. Panel (\ref{fig:Ga6x6mssep}) displays the real and imaginary parts of the eigenvalues, while panel (\ref{fig:Ga6x6mspb}) graphs its arbitrary superposition. In these panels we can straightforwardly see what we have posted above in some figures discussion. That is, the $\kappa_{\mb{\tiny T}}$ parameter leads to a greater mixing between $sh$ (\textcolor{orange}{x}-\textcolor{gdark}{x}) and \textit{lh} (\textcolor{red}{x}-\textcolor{green}{x}), than that for $sh$ (\textcolor{orange}{x}-\textcolor{gdark}{x}) and \textit{hh} (\textcolor{bluel}{x}-\textcolor{blues}{x}). Importantly, it can be observed an interplay of real and imaginary values of $k_{z}$, yielding the $hh$, $lh$ and $sh$ to re-adapt their `perception' of the effective-potential they interact with, whenever the holes travel through re-shaped scattering-profile of QW-acting or QB-acting regions, as the sub-bandmixing augments. Owing to brevity, we present just few evidences. For example, in panel (\ref{fig:Ga6x6mspb}): (i) At $\kappa_{\mb{\tiny T}} = 0.065$ \AA$^{-1}$, the $\Re(k_{z,lh})$ eigenvalues change to $\Im(k_{z,hh})$. (ii) Meanwhile at $\kappa_{\mb{\tiny T}} = 0.135$ \AA$^{-1}$ the $\Im(k_{z,hh})$ eigenvalues is coincident with the $\Im(k_{z,sh})$. In panel (\ref{fig:Ga6x6pb}) we have found similar features, though at slightly lower $\kappa_{\mb{\tiny T}}$ values. For example: (iii) At $\kappa_{\mb{\tiny T}} = 0.065$ \AA$^{-1}$, the $\Re(k_{z,lh})$ eigenvalues evolve into pure $\Im(k_{z,hh})$. (iv) While at $\kappa_{\mb{\tiny T}} = 0.09$ \AA$^{-1}$ arise $\Im(k_{z,sh})$ eigenvalues and the $\Re(k_{z,sh})$ part join the $\Re(k_{z,lh})$ one. From both panels (\ref{fig:Ga6x6mssep}) and (\ref{fig:Ga6x6mspb}), worthwhile underline the $lh-sh$ sub-bandmixing presence for a wider $\kappa_{\mb{\tiny T}}$ scope, on the contrary of that for $hh-sh$, which can  be barely detected within a reduced range of the mixing parameter. Indeed, if we focus on the evolution of the eigenvalues of $lh$ as $\kappa_{\mb{\tiny T}}$ grows, the propagating modes of the $lh$ become alike the $sh$ modes. No doubt this derive from the detected fact, namely: for $k_{z} = 0.09$ \AA$^{-1}$ at $\kappa_{\mb{\tiny T}} = 0.11$ \AA$^{-1}$ it fulfills that $m_{lh}^{*}\approx m_{sh}^{*}\;$, which consequently yields the $lh$ modes turn into $sh$ ones. Panels (\ref{fig:Ga6x6sep}) and (\ref{fig:Ga6x6pb}) plot the same as panels (\ref{fig:Ga6x6mssep})-(\ref{fig:Ga6x6mspb}) but in the absence of $m_{s}$. We have found some slightly-counterintuitive evidences leading us to suppose the inclusion of SO band, as a trigger for modifications of $hh$ spectral distribution. Indeed, the $k_{z,hh}$ (\textcolor{blues}{x}) oscillations range, shows an increment [see panel (\ref{fig:Ga6x6mspb})] respect the interval $\left|\left[0.21 - 0.36\right]\right|$ \AA$^{-1}$ observed without $m_s$. Besides, the $k_{z,hh}$ curve, do not split at the vicinity of $\kappa_{\mb{\tiny T}} = 0.075$ \AA$^{-1}$ as it does in the presence of $m_{s}$. On the other hand, the separation between $k_{z,hh}$ (\textcolor{bluel}{x})  and $k_{z,sh}$ (\textcolor{orange}{x}) at $\kappa_{\mb{\tiny T}} \approx 0.14$ \AA$^{-1}$, vanishes when $m_{s}$ is considered. Thus, the SO-band effect is a robust competitor to the sub-bandmixing, leading the $hh-sh$ interplay to disappear in opposition to what is found when $\kappa_{\mb{\tiny T}}$ grows. Therefore, the inclusion of the term $m_{s}$ represents certain balance in the $lh-sh$ and $hh-sh$ interactions and we remark that a phenomenology of this sort, have been reported before for optical transitions \cite{ekbote,ahn,Park2014,
Kumar2017} and luminescence processes \cite{ekimov1993,singh2016}. Finally, we have retrieved in panels (\ref{fig:Ga4x4sep}) and (\ref{fig:Ga4x4pb}), the $(4 \times 4)$ KL model to compare and to remark several differences. Perhaps the most appealing of them, can be observed for the  $KL_{6 \times 6}$ model, whose sub-bandmixing arises even at the low-mixing regime, while for the  $KL_{4 \times 4}$ case the relevant bandmixing effects take place solely for $\kappa_{\mb{\tiny T}} > 0.09$ \AA$^{-1}$. Besides, there is a clear modification in the ranges and evolution of the $k_{z,hh}$, $k_{z,lh}$ and $k_{z,sh}$ eigenvalues. See for example, in the \textit{root-locus-like} map of panel (\ref{fig:Ga6x6mspb}), the $k_{z,lh}$ branch at the low-mixing regime which starts at $k_{z} = 0.6$ \AA$^{-1}$, while in the $(4 \times 4)$ case, this occurs at $k_{z} = 0.1$ \AA$^{-1}$.

\begin{figure}[tbp]
 \begin{center}
  \subfloat[QW(QB)-acting profiles at top(bottom) panel.]{
  \label{fig:Al6x6mssep}
  \includegraphics[width=0.40\textwidth]{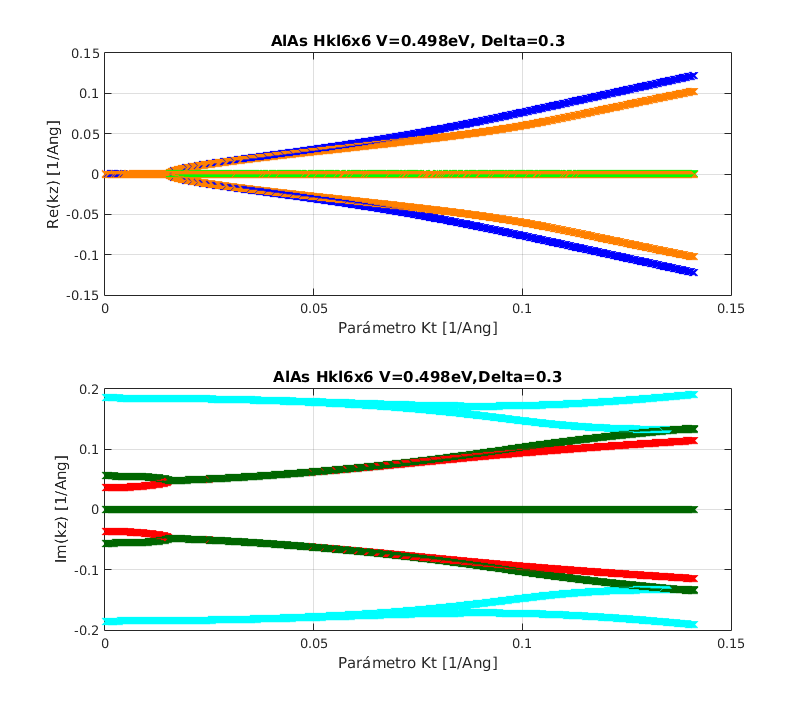}}
    \subfloat[Superposition of the QW(QB)-acting profiles.]{
  \label{fig:Al6x6mspb}
  \includegraphics[width=0.40\textwidth]{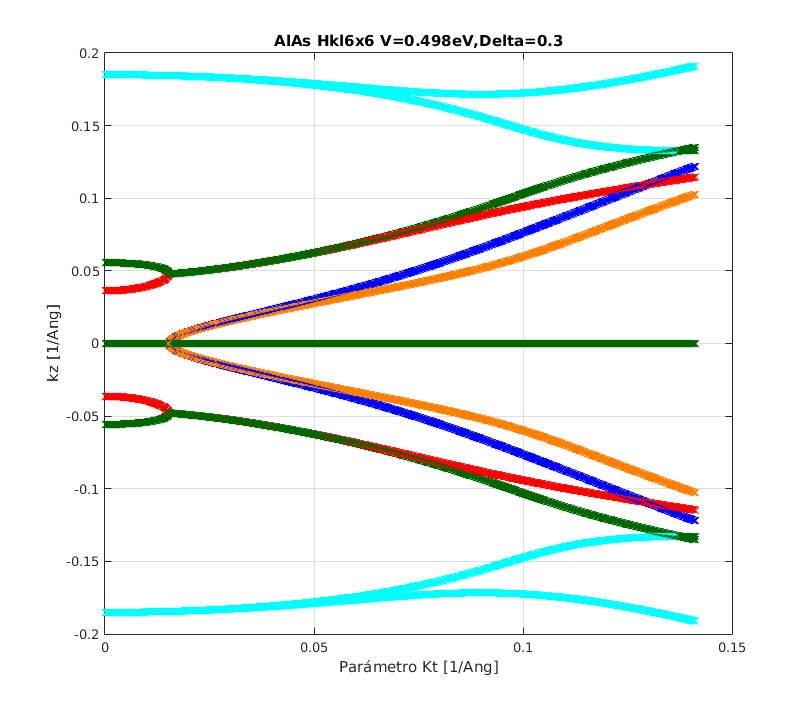}}\\
    \subfloat[QW(QB)-acting profiles at top(bottom) panel.]{
  \label{fig:Al6x6sep}
  \includegraphics[width=0.40\textwidth]{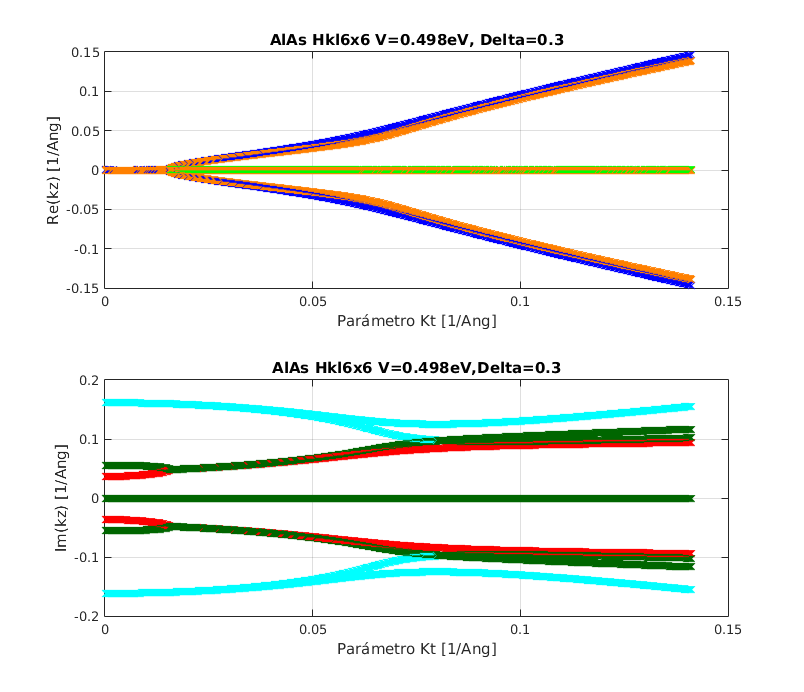}}
    \subfloat[Superposition of the QW(QB)-acting profiles.]{
  \label{fig:Al6x6pb}
  \includegraphics[width=0.40\textwidth]{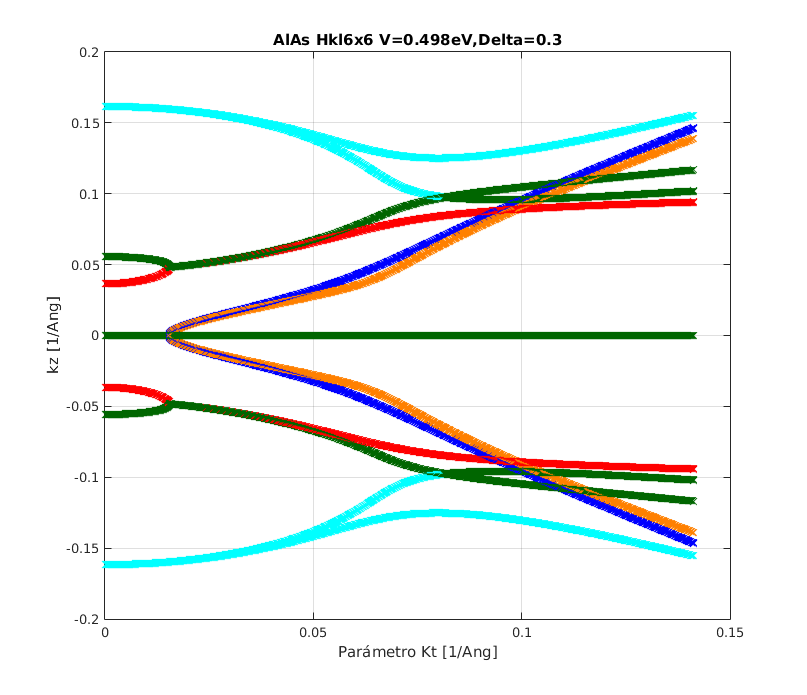}}\\
  \subfloat[QW(QB)-acting profiles at top(bottom) panel.]{
  \label{fig:Al4x4sep}
  \includegraphics[width=0.40\textwidth]{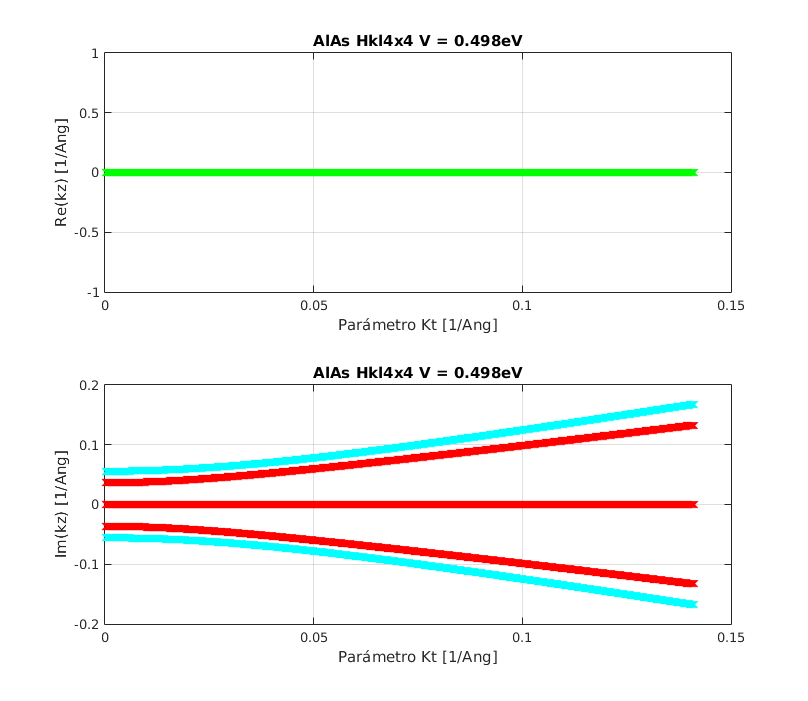}}
  \subfloat[Superposition of the QW(QB)-acting profiles.]{
  \label{fig:Al4x4pb}
  \includegraphics[width=0.40\textwidth]{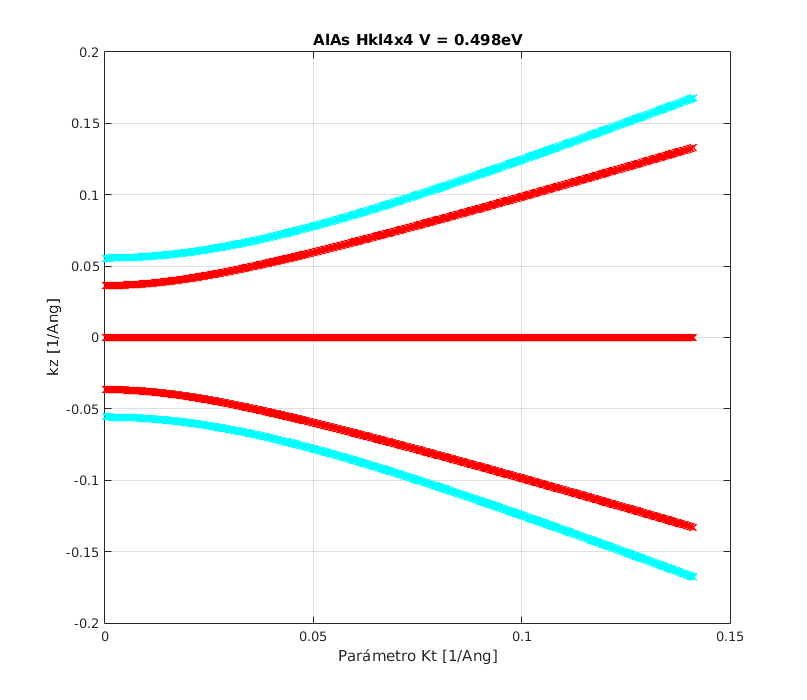}}
 \end{center}
\caption[6vs4al]{\label{AlAspb}
(Color online) Plot of the \textit{root-locus-like} procedure for $k_{z}$ eigenvalues of [$hh$(\textcolor{bluel}{x}-\textcolor{blues}{x}), $lh$(\textcolor{red}{x}-\textcolor{green}{x}), $sh$(\textcolor{orange}{x}-\textcolor{gdark}{x})], as a function of $\kappa_{\mb{\tiny T}}$. We have considered $AlAs$ with $V = 0.498$ eV. For panels III-IV we assumed $m_{s} = 0$. We have taken $\kappa_{\mb{\tiny T}}$  $\in [10^{-6},10^{-1}]$ \AA$^{-1}$.}
\end{figure}

Fig. \ref{AlAspb} plots for $AlAs$, the evolution of the eigenvalues from QEP (\ref{eq:QEP}) as a function of $\kappa_{\mb{\tiny T}}$ taking into account the term $m_{s}$ and also disregarding it. Panel (\ref{fig:Al6x6mssep}) displays the real and imaginary parts of the eigenvalues, while panel (\ref{fig:Al6x6mspb}) graphs its arbitrary superposition for convenience. Similarly as the case of $GaAs$ [see Fig.\ref{GaAspb}], we can observe a larger influence of the SO subband on the $hh$ and $lh$ states, since the hole spectrum is modified considerably in comparison with the case of $KL_{4 \times 4}$, where we only had complex eigenvalues and no mixing between holes in $AlAs$ is detected. However, the subband mixing $sh$(\textcolor{gdark}{x})--$lh$(\textcolor{red}{x}) (at $\kappa_{\mb{\tiny T}} \approx 0.02$\AA$^{-1}$) and $sh$(\textcolor{orange}{x}-\textcolor{gdark}{x})--$hh$(\textcolor{blues}{x}-
\textcolor{bluel}{x}) (at $\kappa_{\mb{\tiny T}} \approx 0.02$ \AA$^{-1}$ and $\kappa_{\mb{\tiny T}} \approx 0.14$ \AA$^{-1}$) occurs at smaller $\kappa_{\mb{\tiny T}}$ in comparison with that of the $GaAs$. We consider this as a consequence of the $m_{s}$ dependence on $\gamma_{2}$, being this semi-empiric parameter larger for $GaAs$ than for $AlAs$. Thereby, there is a more homogeneous $sh-hh$ and $sh-lh$ interplay, since for $GaAs$ we are able to see a greater separation between the hole eigenvalues as $\kappa_{\mb{\tiny T}}$ rises. For example, at $\kappa_{\mb{\tiny T}} \approx 0$, the eigenvalues for $lh$(\textcolor{green}{x}) start at $k_{z} \approx 0.08$\AA$^{-1}$, $sh$ (\textcolor{gdark}{x}-\textcolor{orange}{x}) at $k_{z} = 0$ and $k_{z} \approx 0.18$ \AA$^{-1}$ , and $hh$ (\textcolor{blues}{x}) at $k_{z} \approx 0.22$ \AA$^{-1}$. Notice that when $\kappa_{\mb{\tiny T}}$ increases, the branches of each hole eigenvalue remain mostly with no difference in magnitude between them, until we get the high-mixing regime ($\kappa_{\mb{\tiny T}} \geq 0.1 $\AA$^{-1}$). The $hh$(\textcolor{blues}{x}) eigenstates become an exception at that regime, because their eigenvalues start to increase in magnitude, departing form the other \textit{root-locus} branches. For instance, in the $AlAs$ at $\kappa_{\mb{\tiny T}} = 0$, the $lh$(\textcolor{red}{x}) eigenvalues start at $k_{z} \approx 0.04$ \AA$^{-1}$, for $sh$(\textcolor{gdark}{x}) at $k_{z} \approx 0.055 $\AA$^{-1}$, while the $hh$(\textcolor{bluel}{x}) ones, begin at $k_{z} \approx 1.9 $\AA$^{-1}$. Again,  when $\kappa_{\mb{\tiny T}}$ increases the hole-eigenvalue branches remain very close, but within the high-mixing regime --contrary to the case of $GaAs$--, with the same exception for $hh$ (\textcolor{bluel}{x}), but this time the eigenvalues smoothly decrease in magnitude. By comparing with the ($4 \times 4$) KL model [see panels (\ref{fig:Al4x4sep}) and (\ref{fig:Al4x4pb})], a hole modes modification can be clearly seen and also we have observed a larger mixing-based conversion of the form: $sh \rightarrow hh$, $lh \rightarrow sh$ and $hh \rightarrow sh.\;$ For the $AlAs$, beginning at $\kappa_{\mb{\tiny T}} \geq 0.1 $\AA$^{-1}$, it was observed that the \textit{root-locus} branches for all holes, separate from, each other. It was also confirmed such transitions like $sh$(\textcolor{gdark}{x}) $\rightarrow hh$(\textcolor{bluel}{x}) and consequently at $\kappa_{\mb{\tiny T}} \geq 0.1 $\AA$^{-1}$ it verifies that $k_{z,sh} \equiv k_{z,hh}$, gradually. Furthermore, at the vicinity of $\kappa_{\mb{\tiny T}} \approx 0.14 $\AA$^{-1}$ it fulfills that their evanescent modes are likely the same, \textit{i.e.} $\Im(k_{z,sh}) \approx \Im(k_{z,hh})$. For $GaAs$ we have the opposite, indeed, the  $sh$ (\textcolor{gdark}{x}) modes take apart from those of $hh$(\textcolor{bluel}{x}) at $\kappa_{\mb{\tiny T}} \approx 0.135 $\AA$^{-1}$. The transition $lh$(\textcolor{red}{x}) $\rightarrow sh$(\textcolor{gdark}{x}) is observed within the interval $\kappa_{\mb{\tiny T}} \in \left[0.02,0.06\right] $\AA$^{-1}$, while for higher values the existence of a transfer  $sh$(\textcolor{gdark}{x}) $\rightarrow hh$(\textcolor{blue}{x}), is found. We present in the panels (\ref{fig:Al6x6sep}) and (\ref{fig:Al6x6pb}) the eigenvalue evolution without the term $m_{s}$, to confirm the influence of the SO subband over the $hh$ modes, since the changes in the $hh$ spectrum [mostly seen for $hh$(\textcolor{bluel}{x})] in the presence of $m_{s}$ appear at $\kappa_{\mb{\tiny T}} \geq 0.1 $\AA$^{-1}$, occur now at $\kappa_{\mb{\tiny T}} \approx 0.75 $\AA$^{-1}$ [for $hh$(\textcolor{bluel}{x})]. For $hh$(\textcolor{blues}{x}), it is seen that their eigenvalues are very similar to those of $sh$(\textcolor{orange}{x}) modes. A crossover of this sort, did not happened for the case when $m_{s}$ is taken into account, being the $lh$ and $sh$ spectrum nearly unchanged. In the next subsection we will focus the term $m_{s}$ to get a deeper insight into its physical meaning, as well as its explicit influence on the hole's eigenvalues.

\subsubsection{Profile evolution of the spectral distribution of the QEP}

\begin{figure}[tbp]
 \begin{center}
  \subfloat[Evolution of $k_{z,hh,lh,sh}$ varying $m_{s}$ for GaAs]{
  \label{fig:Ga_compl}
  \includegraphics[width=0.50\textwidth]{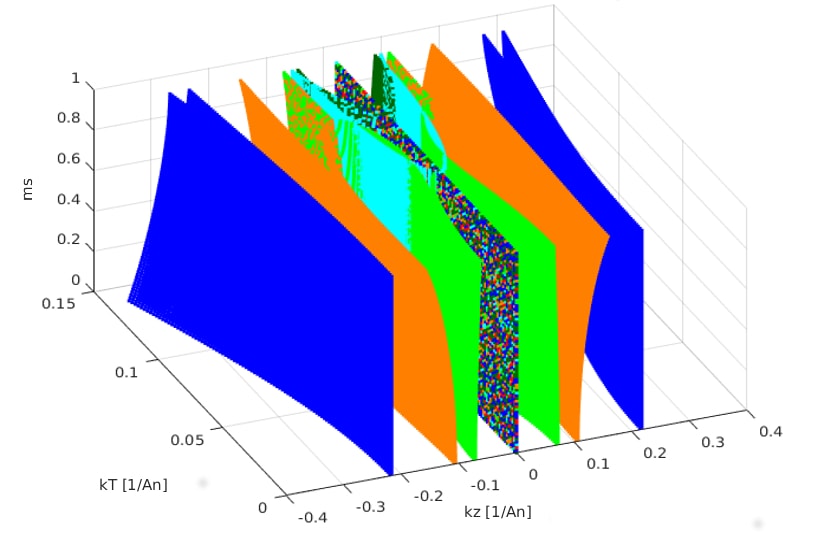}}
    \subfloat[Evolution of $k_{z,hh,lh,sh}$ varying $m_{s}$ for AlAs]{
  \label{fig:Al_compl}
  \includegraphics[width=0.50\textwidth]{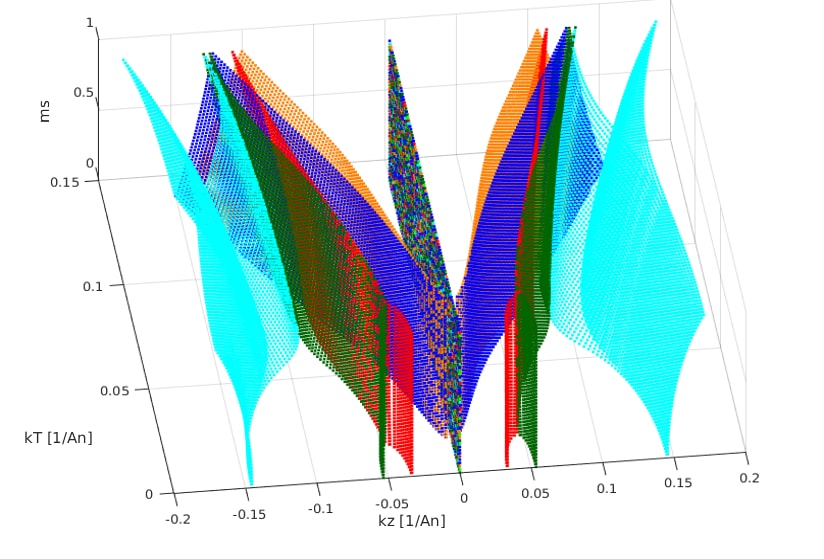}}\\
    \subfloat[Evolution of $k_{z,hh,sh}$ varying $m_{s}$ for GaAs]{
  \label{fig:Ga_hhvshso_compl}
  \includegraphics[width=0.50\textwidth]{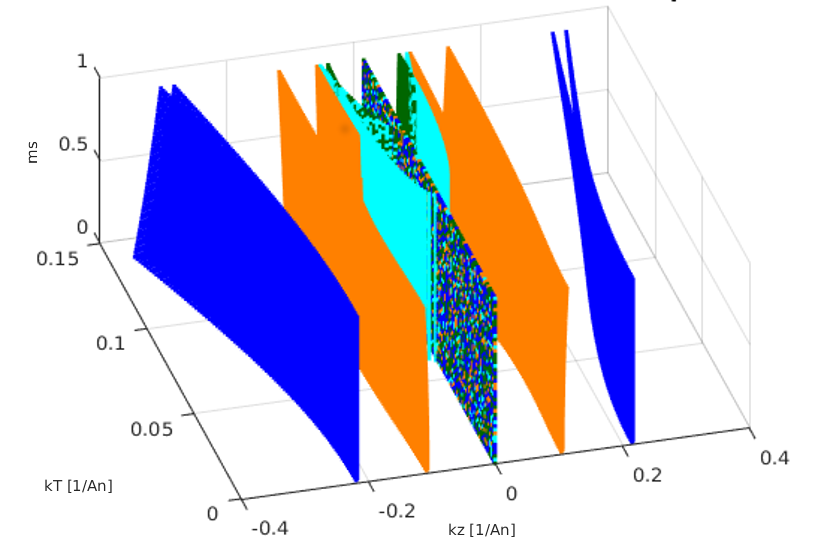}}
    \subfloat[Evolution of $k_{z,hh,sh}$ varying $m_{s}$for AlAs]{
  \label{fig:Al_hhvshso_compl}
  \includegraphics[width=0.50\textwidth]{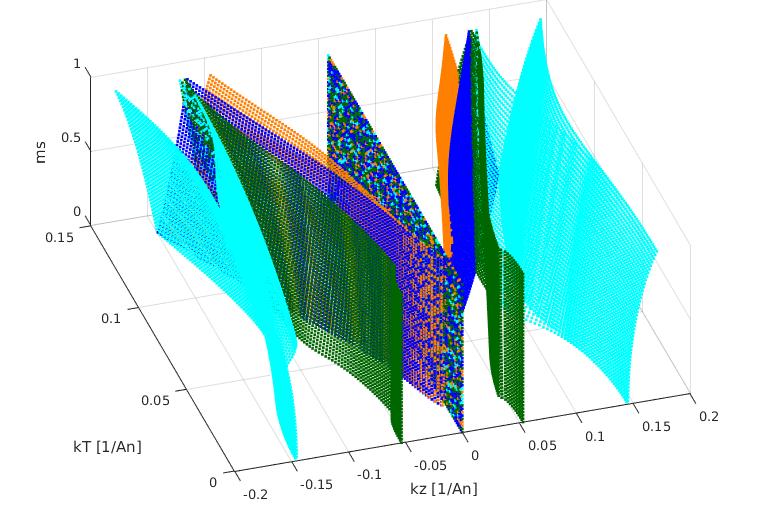}}\\
  \subfloat[Evolution of $k_{z,lh,sh}$ varying $m_{s}$ for GaAs]{
  \label{fig:Ga_lhvshso_compl}
  \includegraphics[width=0.50\textwidth]{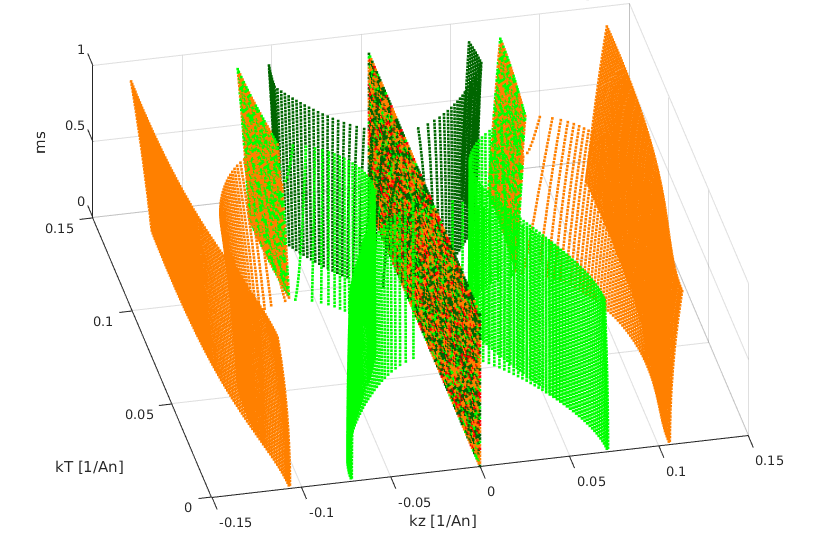}}
  \subfloat[Evolution of $k_{z,lh,sh}$ varying $m_{s}$ for AlAs]{
  \label{fig:Al_lhvshso_compl}
  \includegraphics[width=0.50\textwidth]{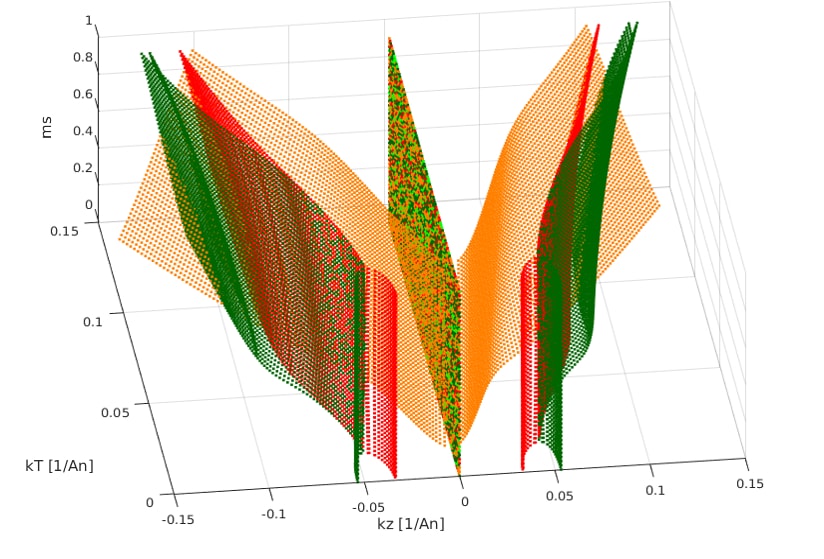}}
 \end{center}
\caption[6vs4al]{\label{ms_compl}[$hh$(\textcolor{bluel}{x}-\textcolor{blues}{x}), $lh$(\textcolor{red}{x}-\textcolor{green}{x}), $sh$(\textcolor{orange}{x}-\textcolor{gdark}{x})]
       $3D$-perspective evolution of the $k_{z} [\AA^{-1}]$ eigenvalues, as a function of the bandmixing parameter $\kappa_{\mb{\tiny T}} [10^{-6},10^{-1}]$ \AA$^{-1}$, showing the profiles for $hh$, $lh$ and $sh$ as a dependence of a $m_{s}$ term percentage.}
\end{figure}

\begin{figure}[tbp]
 \begin{center}
  \subfloat[Plane ($\kappa_{\mb{\tiny T}}k_{z}$) for $GaAs$]{
  \label{fig:Ga_pxy}
  \includegraphics[width=0.40\textwidth]{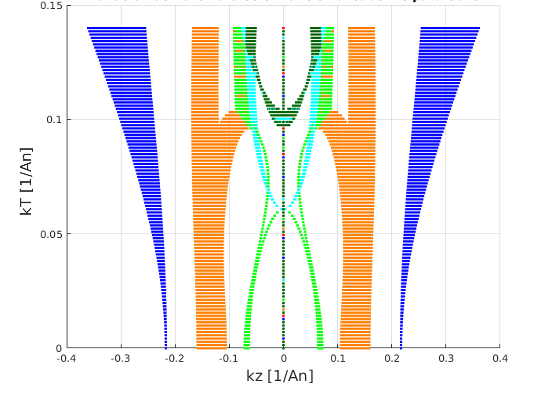}}
    \subfloat[Plane ($\kappa_{\mb{\tiny T}}k_{z}$) for $AlAs$]{
    \label{fig:Al_pxy}
  \includegraphics[width=0.40\textwidth]{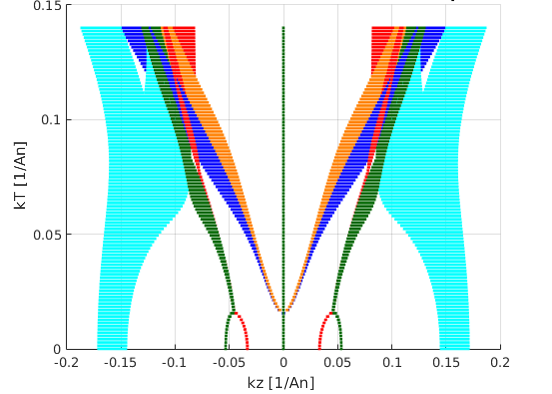}}
 \end{center}
\caption[6vs4al]{\label{ms_pxy}[$hh$(\textcolor{bluel}{x}-\textcolor{blues}{x}), $lh$(\textcolor{red}{x}-\textcolor{green}{x}), $sh$(\textcolor{orange}{x}-\textcolor{gdark}{x})]
$2D$-density map of the [$\kappa_{\mb{\tiny T}}k_{z}$] plane of the eigenvalues evolution as a function of the $m_{s}$ percentage.}
\end{figure}

The Fig.\ref{ms_compl} and Fig.\ref{ms_pxy} display a numerical simulation of how the \textit{sh}-related \textit{off}-diagonal elements $m_{s}$ of the QEP mass-matrix (\ref{eq:B-GEP12}), modify the hole eigenvalues spectrum.

In Fig. \ref{ms_compl} the $3D$-perspective profiles of the $k_{z,hh,lh,sh}$ eigenvalues are shown by varying a percentage of the term $m_{s}$ [0(0\%), 1(100\%)] and $\kappa_{\mb{\tiny T}}$. Fig. \ref{ms_pxy} presents a $2D$-density map of Fig. \ref{ms_compl}, projected on the [$\kappa_{\mb{\tiny T}}k_{z}$] plane. The phenomenology for $GaAs$ can be observed in the graphs: \ref{fig:Ga_compl}, \ref{fig:Ga_hhvshso_compl} and \ref{fig:Ga_lhvshso_compl}, while for $AlAs$ we display the panels:  \ref{fig:Al_compl}, \ref{fig:Al_hhvshso_compl} and \ref{fig:Al_lhvshso_compl}. As mentioned above in the subsection \ref{subsec:profiles}, it can be observed in Fig. \ref{fig:Ga_compl}, the variation of the ranges for the $k_{z,hh,lh,sh}$ eigenvalues by taking the term $m_s$. One can see, for example, the notable behavior for $k_{z,hh}$(\textcolor{blues}{x}), who's variation range goes in the interval $\left|\left[0.21,0.36\right]\right|$ \AA$^{-1}$. Fig. \ref{fig:Ga_hhvshso_compl} exhibits the changes for the $k_{z,sh}$(\textcolor{orange}{x}-\textcolor{gdark}{x}) and $k_{z,hh}$(\textcolor{blues}{x}-\textcolor{bluel}{x}) eigenvalues, being this last case mostly negligible with $m_{s}$ increasing, near $\kappa_{\mb{\tiny T}} \approx 0$, though at the high-mixing regime the spectrum variation  turns more considerable. As it can be straightforwardly seen for $\kappa_{\mb{\tiny T}} \approx 0.1$ \AA$^{-1}$ and within the interval $\kappa_{\mb{\tiny T}} \in \left|\left[0.24,0.3\right]\right|$ \AA$^{-1}$ when $m_{s} \in \left[0,1\right]$, the $k_{z,hh}$(\textcolor{blues}{x}) spectral distribution approaches to that of $k_{z,sh}$(\textcolor{orange}{x}). Besides, $k_{z,hh}$(\textcolor{bluel}{x}) and $k_{z,sh}$(\textcolor{gdark}{x}) spectrums explicitly cross at $\kappa_{\mb{\tiny T}} \approx 0.1$ \AA$^{-1}$. These features can be more accurately observed in the [$\kappa_{\mb{\tiny T}}k_{z}$] plane of Fig. \ref{fig:Ga_pxy}. Furthermore, we found in Fig. \ref{fig:Ga_lhvshso_compl} that the eigenvalues of $k_{z,lh}$(\textcolor{green}{x}) --remaining nearly unchanged at $\kappa_{\mb{\tiny T}} \leq 0.1$ \AA$^{-1}$--, can evolve for higher mixing values, yielding to slightly match those of the $k_{z,sh}$(\textcolor{orange}{x}). At this point, it must be remarked firstly: that the $k_{z,hh}$ eigenvalues variation with $m_{s}$, is the largest comparing with that of $k_{z,lh,sh}$. Secondly: the off-diagonal term $m_{s}$ influences over the $hh-lh-sh$ spectral properties and propagating modes.

In Fig. \ref{fig:Al_compl} we can observe the influence of the term $m_{s}$ over the $k_{z,hh,lh,sh}$ eigenvalues. It is not difficult to note, the considerable scope for $k_{z,hh}$(\textcolor{bluel}{x}) variations under zero-mixing regime ($\kappa_{\mb{\tiny T}} = 0$). This behavior differs from that of $GaAs$ discussed above, where under the same condition the $k_{z,hh}$(\textcolor{orange}{x}) spectrum, were the one that changes. This is likely because $\Delta_{0}^{GaAs} > \Delta_{0}^{AlAs}$ and $\gamma_{2}^{GaAs} > \gamma_{2}^{AlAs}$, leading therefore to the observed effects on $k_{z,hh}$(\textcolor{blues}{x}-\textcolor{bluel}{x}) under the high-mixing regime [see Fig.\ref{fig:Ga_hhvshso_compl}], as well as to the discussed behavior of $k_{z,hh}$(\textcolor{bluel}{x}) at $\kappa_{\mb{\tiny T}} = 0$ [see Fig.\ref{fig:Ga_compl}]. On the other hand, as $\kappa_{\mb{\tiny T}}$ grows, an appealing interplay between propagating and evanescent modes rises. See for example the region surrounding $\kappa_{\mb{\tiny T}} \approx 0.02$ \AA$^{-1}$, where (at any fraction of $m_s$) there are transitions such as:  $lh$(\textcolor{red}{x}) $\rightarrow$ $sh$(\textcolor{gdark}{x}) for evanescent modes [see Fig. \ref{fig:Al_lhvshso_compl}], and $hh$(\textcolor{blues}{x}) $\rightarrow$ $sh$(\textcolor{orange}{x}) for propaganting modes [see Fig.\ref{fig:Al_hhvshso_compl}]. Correspondingly, but at high-mixing regime the transitions $sh$(\textcolor{gdark}{x}) $\rightarrow$ $hh$(\textcolor{bluel}{x}) were found at the interval [$0.75 \leq \kappa_{\mb{\tiny T}} \geq 0.14$] \AA$^{-1}$, depending on the fraction of $m_{s}$ that have been taken into account. We have also explored the $k_{z,sh}$(\textcolor{orange}{x}) spectral distribution at high-mixing regime and higher(lower) values were obtained without(with) $m_{s}$, whenever [$0.1 \leq k_{z,sh}(\textcolor{orange}{x}) \geq 0.14$] \AA$^{-1}$. Finally, by disregarding $m_s$ at $\kappa_{\mb{\tiny T}} \approx 0.1$ \AA$^{-1}$ we have obtained that $k_{z,sh}$(\textcolor{orange}{x})$ = k_{z,lh}$(\textcolor{red}{x}), meanwhile at $\kappa_{\mb{\tiny T}} \approx 0.12$ \AA$^{-1}$ it verifies that $k_{z,sh}$(\textcolor{orange}{x})$ = k_{z,sh}$(\textcolor{gdark}{x}). This last evidence, opposes to the case with $m_{s}$ where the $k_{z,sh}$(\textcolor{orange}{x}) eigenvalues only separate from the $k_{z,hh}$(\textcolor{blues}{x}) when $\kappa_{\mb{\tiny T}} \geq 0.05$ \AA$^{-1}$ [see in Fig.\ref{fig:Al_pxy}].

\section{Concluding Remarks}
\label{sec:Con}

We have presented a numeric-computational procedure to analyze the spectral distribution for the eigenvalues of $N$-coupled components QEP. The sufficient conditions to solve the correlated GEP have been redefined and the examination of the limit case was evaluated by fulfilling the requirements imposed previously \cite{mendoza,guillermo}. We conclude that the GEP solution, based on a Simultaneous Triangularization scheme derived elsewhere \cite{mendoza,guillermo} for $N$-coupled systems is possible, if and only if, the off-diagonal elements of the QEP-mass matrix are zero or can be disregarded. The last is not the case for de extended $KL_{6 \time 6}$ model exercised here, thus we proceed successfully with a Generalized Schur Decomposition method.

It is worthy to emphasize that without any doubt, the $sh-hh-lh$ interplay mechanism, is reliable with the presence of $m_{s}$ as off-diagonal elements in the QEP mass-matrix (\ref{eq:B-GEP12}). In this concern, the active influence by these \textit{sh}-related terms over the $hh-lh$ spectral distribution, does not seem to be susceptible to be disregarded or even zeroed as had been assumed elsewhere \cite{ekbote,harrison2005}. On the contrary, we suggest this SO-like act upon the hole spectrum variance and the $sh-hh-lh$ interplay, \emph{via} the $sh$-related terms $m_{s}$, as a strong competitor of the standardized bandmixing influence performed by the $\kappa_{\mb{\tiny T}}$ parameter.  Such effect depends on the SO-subband gap $\Delta_{0}$ and on the L\"{u}ttinger parameter $\gamma_{2}$, which is proportional to $m_s$.

\bibliographystyle{elsarticle-num}
\bibliography{biblio7a}

\appendix
\section{Matrix elements of $H_{\mb{\tiny KL}(6 \times 6)}$}
\label{app:KL6}

For the Hamiltonian (\ref{eq:KL6-1}), we have
\begin{flalign}
 \label{eq:HKL6-ap1}
 P &=\frac{\hbar^{2}}{2m}\text{$\gamma_{1}$} \left(\text{$k_{x}^{2}+k_{y}^{2}$}+\text{$k_{z}$}^2\right);\\
 \nonumber
 Q &=\frac{\hbar^{2}}{2m}\text{$\gamma_{2}$} \left(\text{$k_{x}^{2}+k_{y}^{2}$}-2 \text{$k_{z}$}^2\right);\\
 \nonumber
 S &=\sqrt{3}\frac{\hbar^{2}}{m} \text{$\gamma_{3}$} \text{$k_{z}$} (\text{$k_{x}$}-i \text{$k_{y}$});\\
 R &= \frac{\hbar^{2}}{2m} \sqrt{3} \left(\gamma_{2}(\text{$k_{y}$}^2-\text{$k_{x}$}^2)+2 i \text{$\gamma_{3}$} \text{$k_{x}$} \text{$k_{y}$} \right),
 \nonumber
\end{flalign}
\noindent while for its modified expression (\ref{eq:KL6-2}) we use
\begin{flalign}
 \label{eq:HKL6-ap2}
 A_{1}&=\frac{\hbar^{2}}{2m}(\gamma_{1}+\gamma_{2});\\
 \nonumber
 A_{2} &= \frac{\hbar^{2}}{2m}(\gamma_{1}-\gamma_{2});\\
 \nonumber
 B_{1} &= \frac{\hbar^{2}}{2m}(\gamma_{1}+2\gamma_{2});\\
 \nonumber
 B_{2} &= \frac{\hbar^{2}}{2m}(\gamma_{1}-2\gamma_{2});\\
 \nonumber
 H_{11} &= A_{1}(\text{$k_{x}$}^2+k_{y}^2) -B_{2}\frac{\partial^{2}}{\partial z^{2}};\\
 \nonumber
 H_{12} &= \frac{\hbar^{2}}{2m} \sqrt{3} \left(\gamma_{2}(\text{$k_{y}$}^2-\text{$k_{x}$}^2)+2 i \text{$\gamma_{3}$} \text{$k_{x}$} \text{$k_{y}$} \right);\\
 \nonumber
 H_{13} &= \frac{\hbar^{2}}{2m}\sqrt{6}\left(\gamma_{2}(\text{$k_{y}$}^2-\text{$k_{x}$}^2)+2 i \text{$\gamma_{3}$} \text{$k_{x}$} \text{$k_{y}$} \right);\\
 \nonumber
 H_{14} &= i\frac{\hbar^{2}}{2m}\sqrt{6}\text{$\gamma_{3}$}(\text{$k_{x}$}-i \text{$k_{y}$})\frac{\partial}{\partial z};\\
 \nonumber
 H_{15} &= i\frac{\hbar^{2}}{2m} \sqrt{3}\text{$\gamma_{3}$} (\text{$k_{x}$}-i \text{$k_{y}$})\frac{\partial}{\partial z};\\
 \nonumber
 H_{22} &= A_{2}(\text{$k_{x}$}^2+k_{y}^{2}) -B_{1}\frac{\partial^{2}}{\partial z^{2}};\\
 \nonumber
 H_{23} &= \frac{\hbar^{2}}{2m}\sqrt{2}\text{$\gamma_{2}$} \left(\text{$k_{x}^{2}+k_{y}^{2}$}+2 \frac{\partial^{2}}{\partial z^{2}}\right);\\
 \nonumber
 H_{24} &= -i\frac{\hbar^{2}}{2m}3\sqrt{2}\text{$\gamma_{3}$}(\text{$k_{x}$}-i \text{$k_{y}$})\frac{\partial}{\partial z};\\
 \nonumber
 H_{33} &= \frac{\hbar^{2}}{2m}\text{$\gamma_{1}$} \left(\text{$k_{x}^{2}+k_{y}^{2}$}-\frac{\partial^{2}}{\partial z^{2}}\right)+\Delta_{0}.&&
\end{flalign}

Here $\gamma_{1},\gamma_{2}$ and $\gamma_{3}$ represent the semi-empirical L\"uttinger parameters and $\Delta_{0}$ represents the spin orbit band gap energy. In the expression (\ref{eq:KL6-2}), the states taken into account are the ones representing the interactions of the holes \textit{$hh_{+3/2}$}, \textit{$lh_{-1/2}$}, \textit{$lh_{+1/2}$}, \textit{$hh_{-3/2}$} and
$sh$.

\section{QEP of $H_{\mb{\tiny KL}(6 \times 6)}$}
 \label{app:QEP6}

For the QEP matrices (\ref{eq:C-QEP6})-(\ref{eq:K-QEP6}), we have defined
\begin{flalign}
 \label{eq:QEP-ap1}
  \mathcal{H}_{14}=-\frac{\hbar^{2}}{2m}\sqrt{6}\gamma_{3}(k_{x}-ik_{y});\\
  \nonumber
  \mathcal{H}_{15}=-\frac{\hbar^{2}}{2m}\sqrt{3}\gamma_{3}(k_{x}-ik_{y});\\
  \nonumber
  \mathcal{H}_{24}=\frac{\hbar^{2}}{2m}3\sqrt{2}\gamma_{3}(k_{x}-ik_{y});\\
  \nonumber
  \mathcal{H}_{33}=\frac{\hbar^{2}}{2m}\gamma_{1}\kappa_{\mb{\tiny T}}^{2}+\Delta_{0}+V(z)-E;\\
  \nonumber
  m_{1}=m_{6}=B_{2};\\
  \nonumber
  m_{2}=m_{5}=B_{1};\\
  \nonumber
  m_{3}=m_{4}=\frac{\hbar^{2}}{2m}\gamma_{1};\\
  \nonumber
  m_{s}=-\frac{\hbar^{2}}{2m}2\sqrt{2}\gamma_{2};\\
  \nonumber
  a_{1}=A_{1}\kappa_{\mb{\tiny T}}^{2}+V(z)-E;\\
  \nonumber
  a_{2}=A_{2}\kappa_{\mb{\tiny T}}^{2}+V(z)-E;\\
  \nonumber
  a_{s}=\frac{\hbar^{2}}{2m}\sqrt{2}\gamma_{2}\kappa_{\mb{\tiny T}}^{2};
\end{flalign}

Here $V(z)$ and $E$ represent the scattering potential and the incident energy of the charge carriers, respectively

\section{Validation of sufficient conditions and STR of GEP}
 \label{app:NecSuf}

\begin{figure} [ht]
 \begin{center}
 \subfloat[Evaluation of the condition (\ref{cond2}) for $GaAs$.]{
 \label{fig:Ga_3d_sms}
   \includegraphics[height = 0.20\textheight]{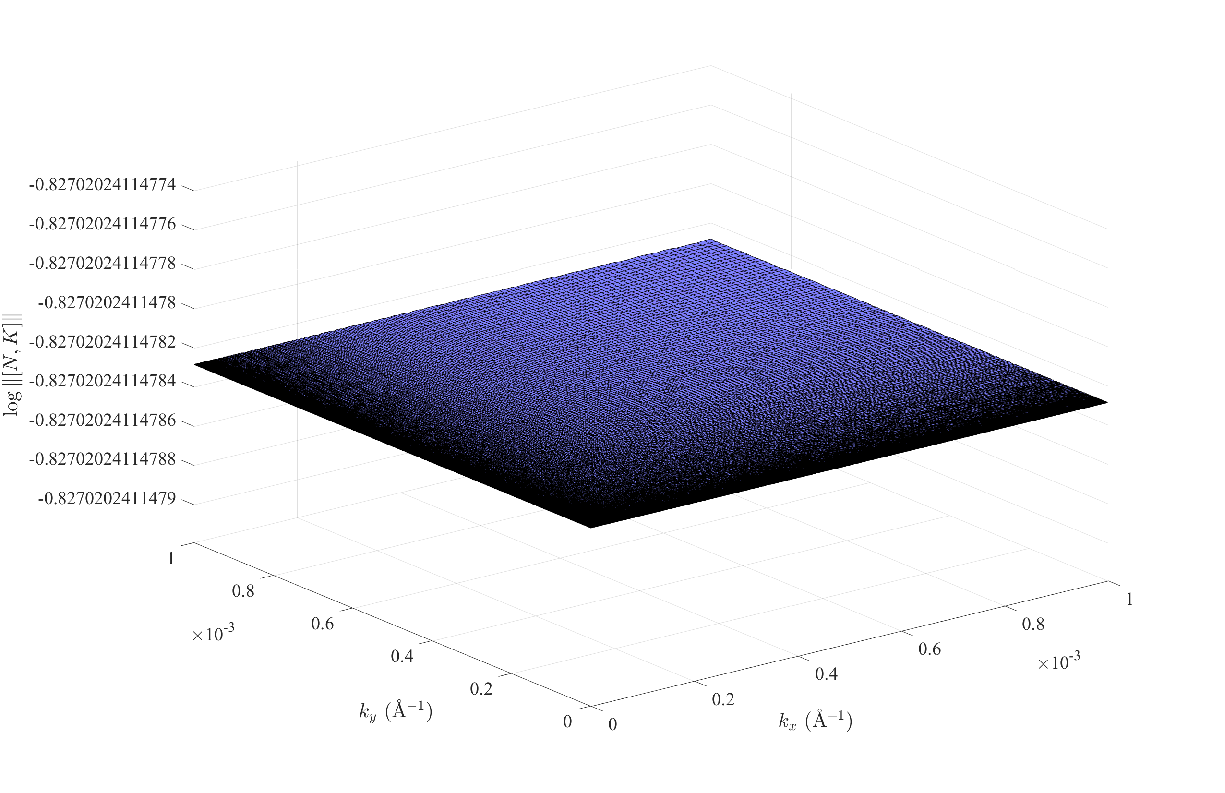}}
  \subfloat[Evaluation of the condition (\ref{cond2}) for $AlAs$.]{
  \label{fig:Al_3d_sms}
   \includegraphics[height = 0.20\textheight]{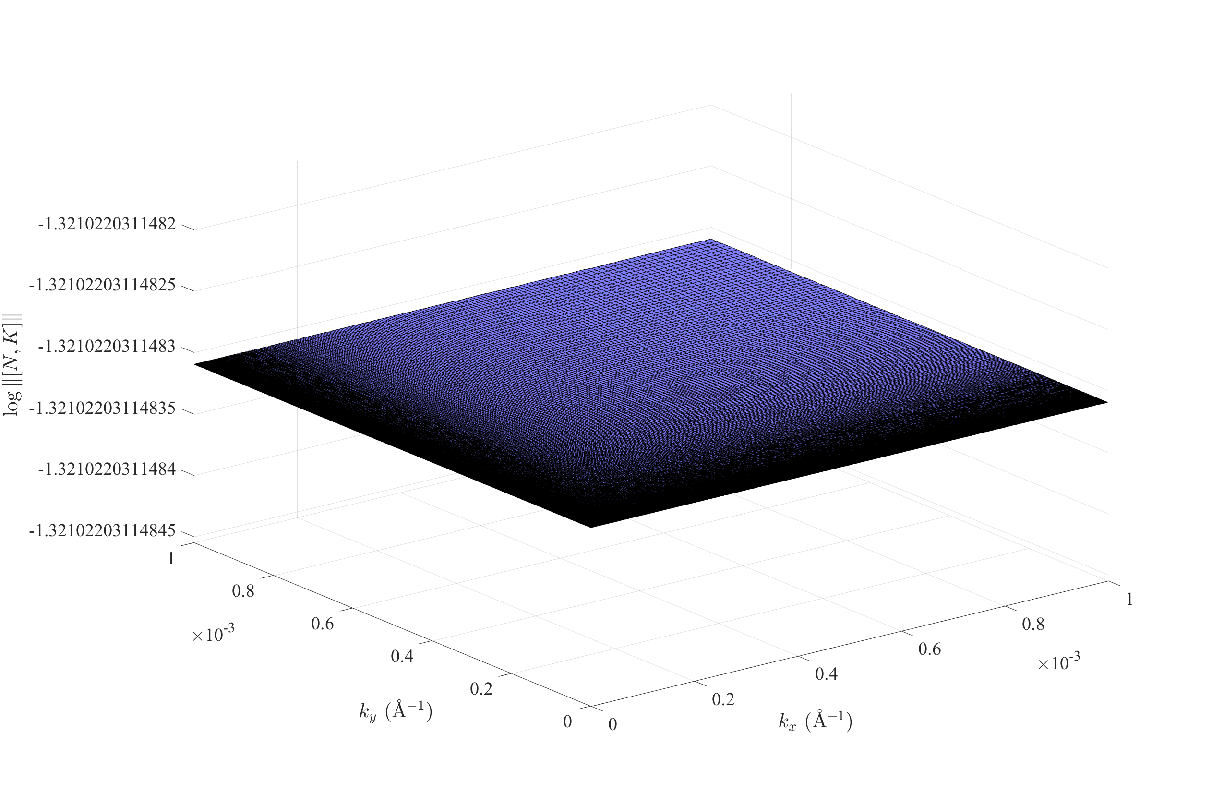}}
\end{center}
\caption[graf]
   {\label{con_sms}(Color online) Spectral norm of the condition (\ref{cond2}) for $GaAs$ [panel (I)] and $AlAs$ [panel (II)] taking $\kappa_{\mb{\tiny T}} \in [10^{-5},10^{-3}]$.}
\end{figure}

\begin{figure} [ht]
 \begin{center}
 \subfloat[Evaluation of the condition (\ref{cond1}) for $GaAs$.]{
 \label{fig:Ga_3d_sms}
   \includegraphics[height = 0.25\textheight]{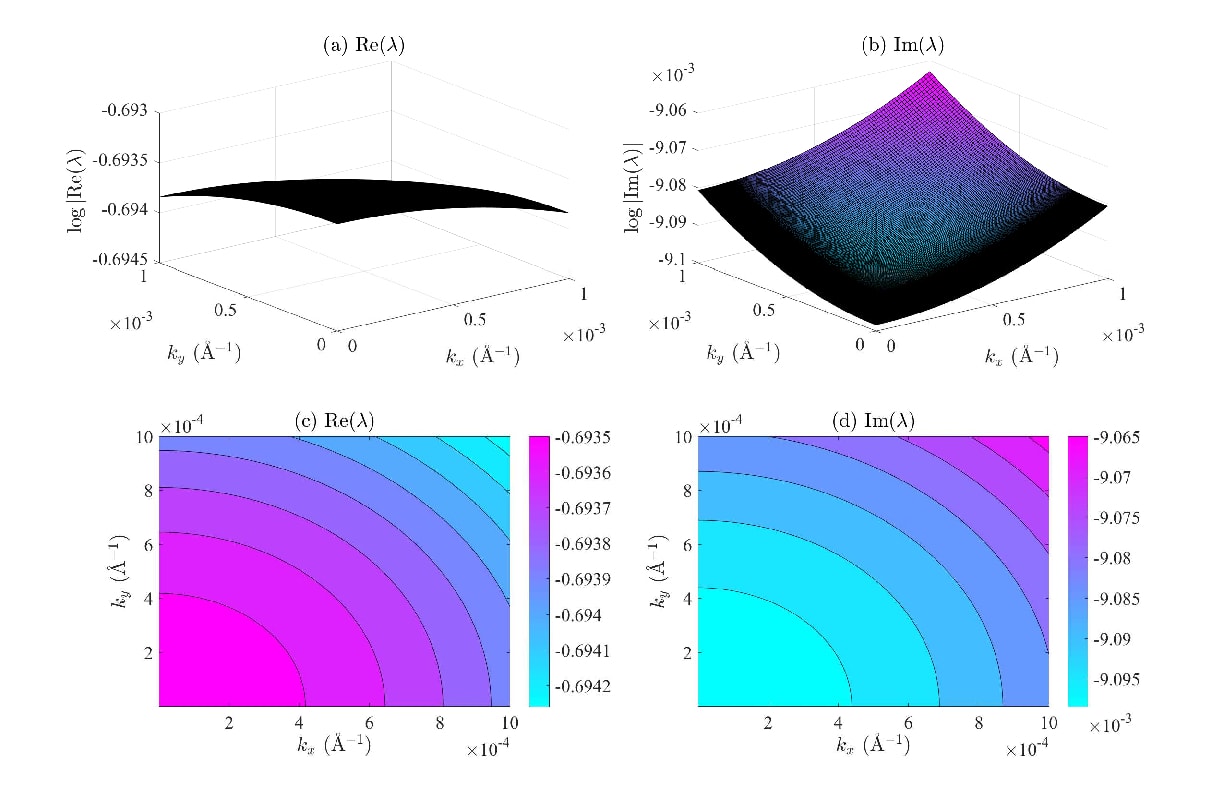}}\\
    \subfloat[Evaluation of the condition (\ref{cond1}) for $AlAs$.]{
\label{fig:Al_3d_sms}
   \includegraphics[height = 0.25\textheight]{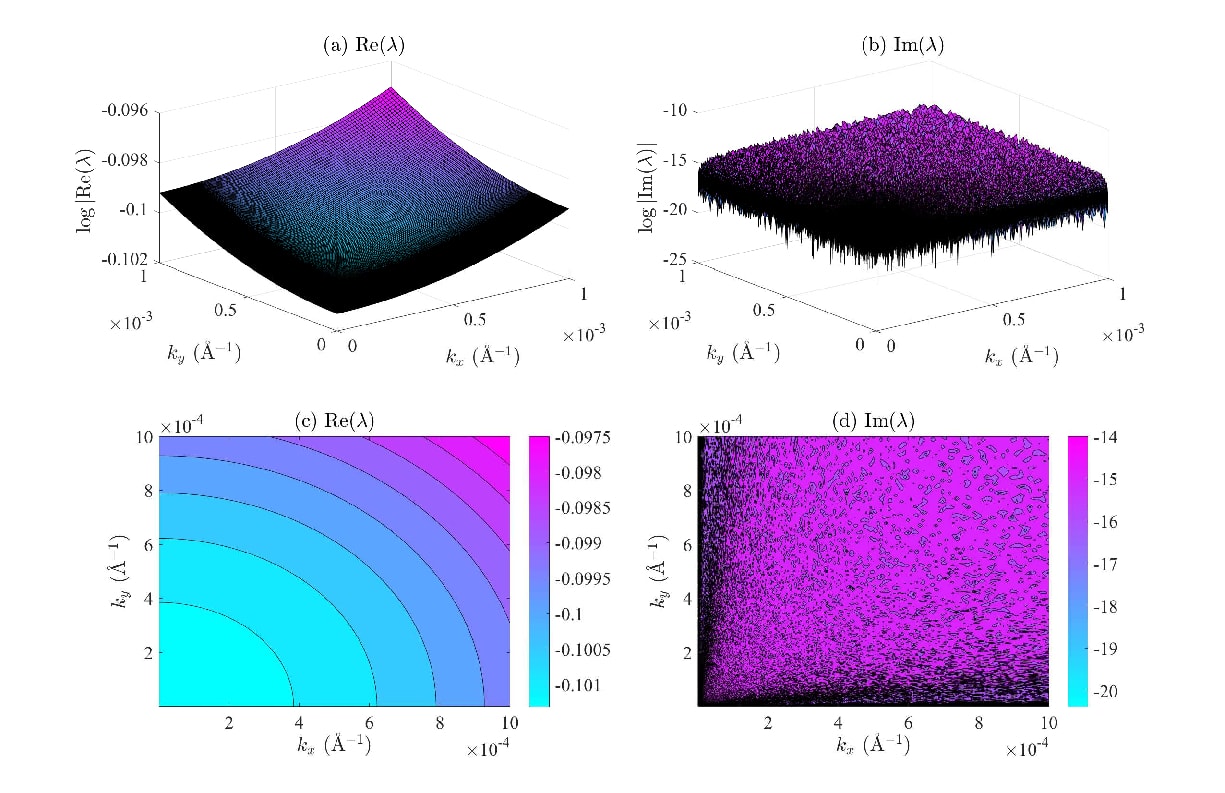}}
\end{center}
\caption[graf]
   {\label{3d_sms} (Color online) Block (a)/(b) maps the spectral distribution of the condition (\ref{cond1}) and validation of the STR associated to GEP (\ref{eq:QEP-Lin2}) for $GaAs/AlAs$. Each inside-block panel (a)/(b), shows the $3D$-contours for the Re($\lambda$)/Im($\lambda$) parts as function of the components $k_{x}$ and $k_{y}$. Meanwhile, each inside-block panel (c)/(d) displays density maps for the Re($\lambda$)/Im($\lambda$) parts. The parameter $\kappa_{\mb{\tiny T}}$ was taken in the range [$10^{-5},10^{-3}$].}
\end{figure}

\begin{figure} [tbp]
 \begin{center}
 \subfloat[Evaluation of (\ref{cond2}) for $GaAs$ without $m_s$.]{
\label{fig:Ga_3d_ms}
   \includegraphics[height = 0.15\textheight]{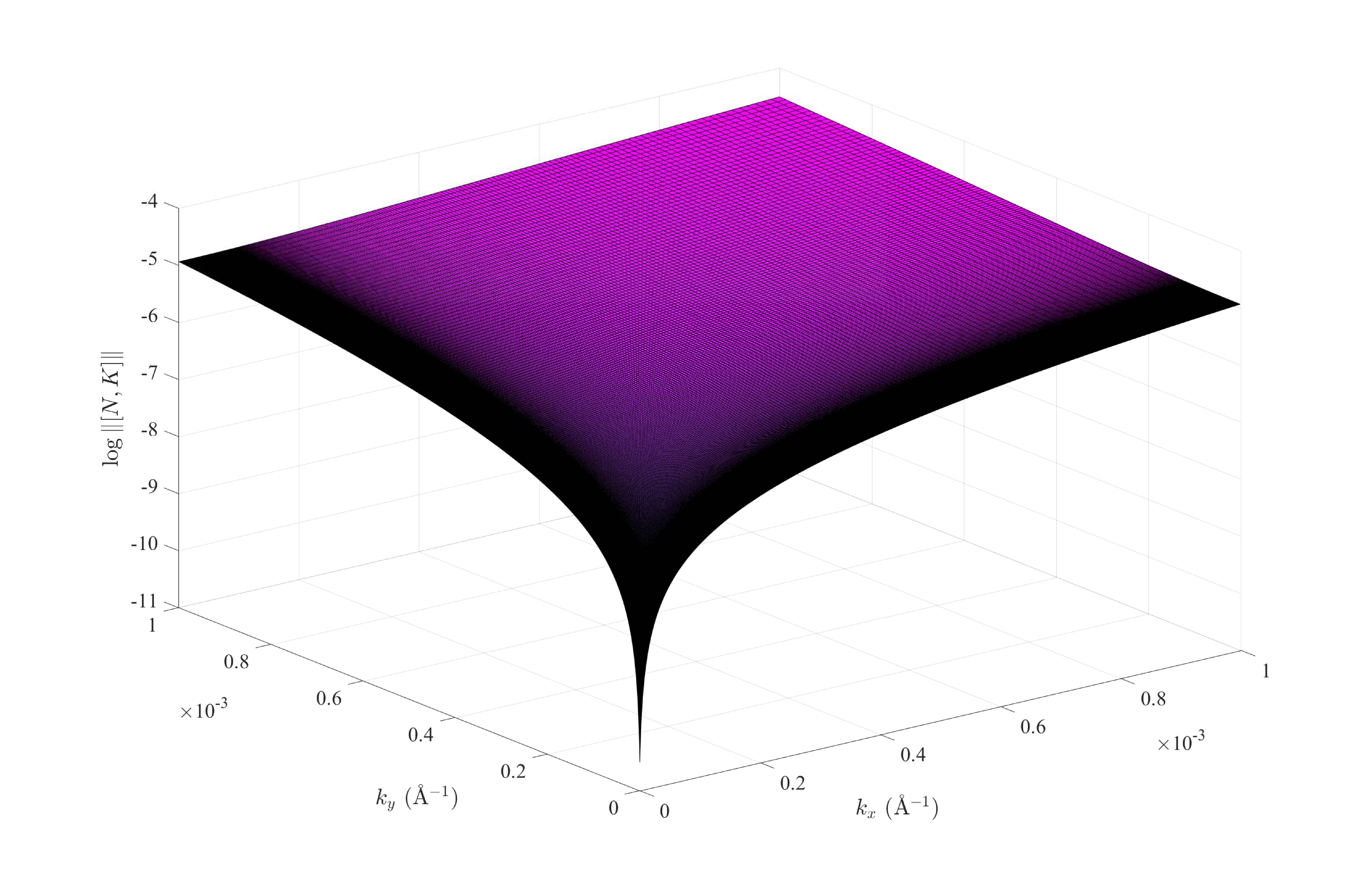}}
    \subfloat[Evaluation of (\ref{cond2}) for $AlAs$ without $m_s$.]{
\label{fig:Al_3d_ms}
   \includegraphics[height = 0.15\textheight]{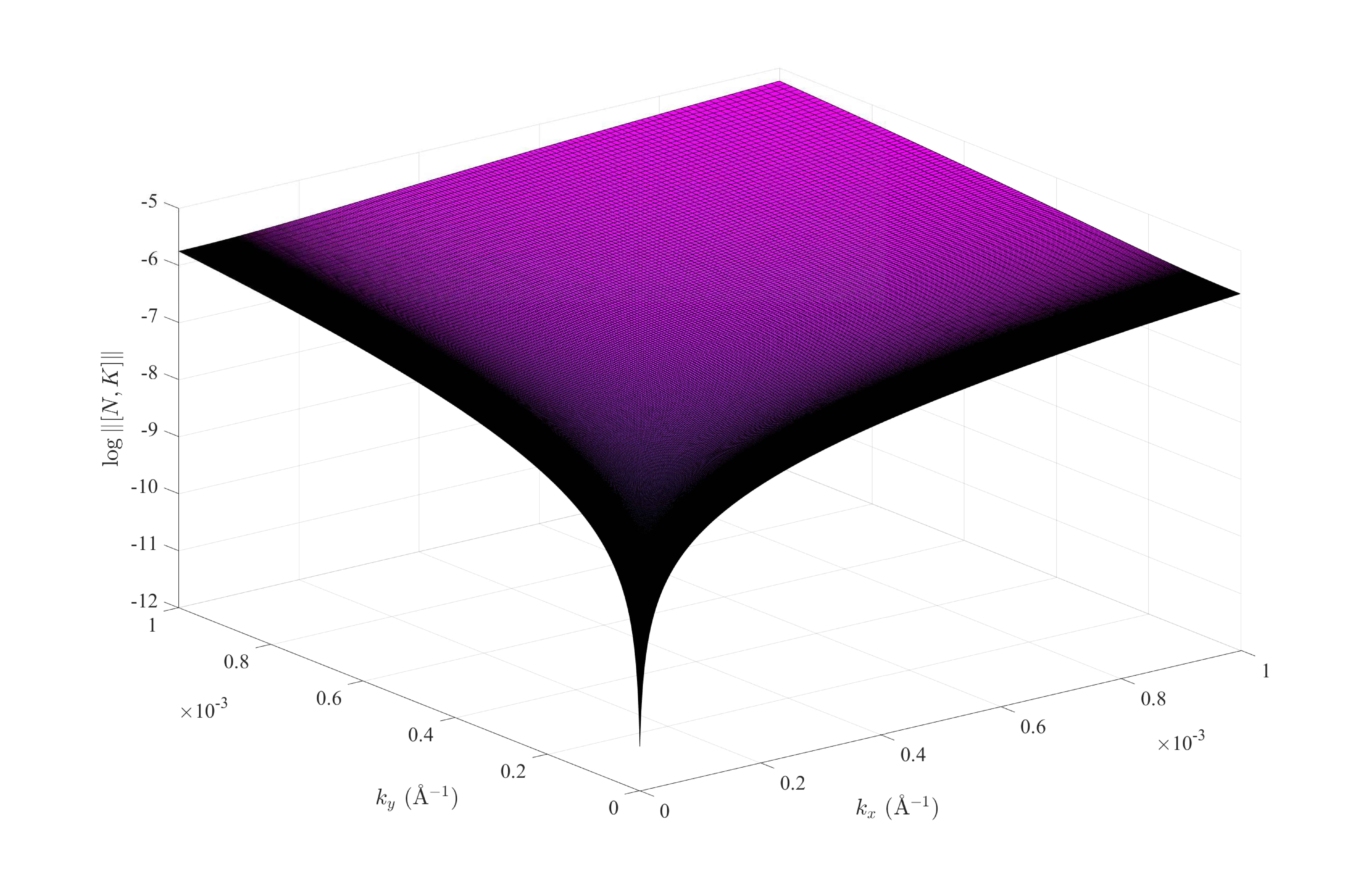}}
\end{center}
\caption[graf]
   {\label{con_ms} (Color online) Spectral norm simulation of the condition (\ref{cond2}) for $GaAs$ [panel (I)] and $AlAs$ [panel (II)], where $\kappa_{\mb{\tiny T}} \in [10^{-5},10^{-3}]$ \AA$^{-1}$.}
\end{figure}

\begin{figure} [ht]
 \begin{center}
 \subfloat[Evaluation of the condition (\ref{cond1}) for $GaAs$ without the term $m_{s}$.]{
  \label{fig:Ga_3d_ms}
   \includegraphics[height = 0.25\textheight]{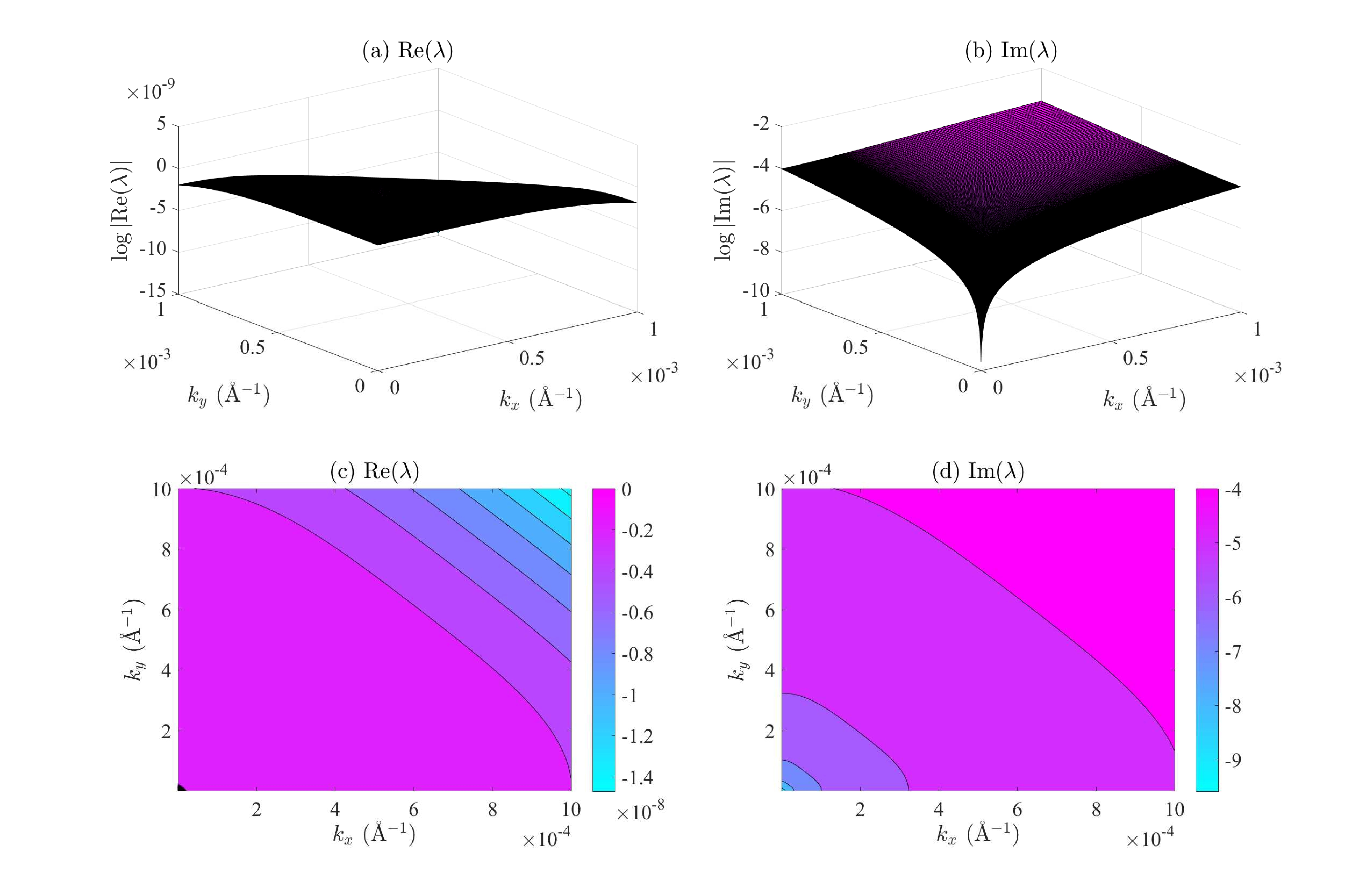}}\\
    \subfloat[Evaluation of the condition (\ref{cond1}) for $AlAs$ without the term $m_{s}$.]{
\label{fig:Al_3d_ms}
   \includegraphics[height = 0.25\textheight]{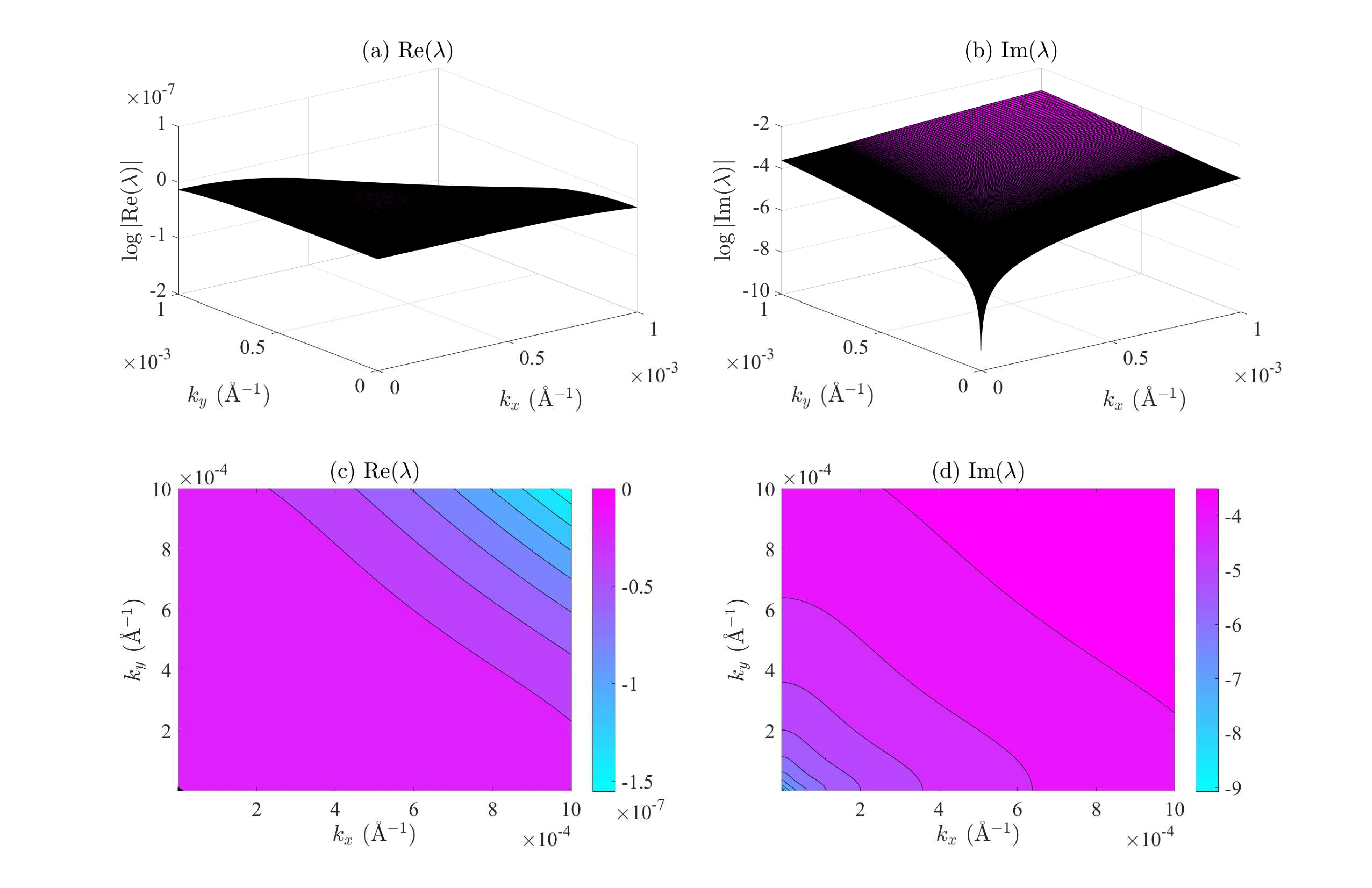}}
\end{center}
\caption[graf]
   {\label{3d_ms}(Color online) Spectral distribution of the condition (\ref{cond1}) for validating the STR of the associated GEP for $GaAs$ and $AlAs$. Each inside-block panel (a)/(b) plots the $3D$-contours for the Re($\lambda$)/Im($\lambda$) parts as function of the components $k_{x}$ and $k_{y}$. While, each inside-block panel (c)/(d) shows density maps for the Re($\lambda$)/Im($\lambda$) parts. The parameter $\kappa_{\mb{\tiny T}}$ was taken in the interval [$10^{-5}, 10^{-3}$] \AA$^{-1}$.}
\end{figure}

\end{document}